\documentclass[twocolumn,twocolappendix,usenames,dvipsnames]{aastex631}

\usepackage{comment}

\begin{document}

\title{Velocity Structure Correlations between the Nebular,  Molecular, and Atmospheric Gases in the Cores of Four Cool Core Clusters}

\author{Muzi Li}
\affiliation{Department of Physics and Astronomy, University of Waterloo, Waterloo, ON N2L 3G1, Canada}
\affiliation{Waterloo Centre for Astrophysics, Waterloo, ON N2L 3G1, Canada}

\author{B.R. McNamara}
\affiliation{Department of Physics and Astronomy, University of Waterloo, Waterloo, ON N2L 3G1, Canada}
\affiliation{Waterloo Centre for Astrophysics, Waterloo, ON N2L 3G1, Canada}

\author{Alison L. Coil}
\affiliation{Department of Astronomy and Astrophysics, University of California, San Diego, La Jolla, CA 92093, USA}

\author{Marie-Jo\"{e}lle Gingras}
\affiliation{Department of Physics and Astronomy, University of Waterloo, Waterloo, ON N2L 3G1, Canada}
\affiliation{Waterloo Centre for Astrophysics, Waterloo, ON N2L 3G1, Canada}

\author{Fabrizio Brighenti}
\affiliation{Dipartimento di Fisica e Astronomia, Università di Bologna, Via Gobetti 93/2, 40122, Bologna, Italy}

\author{H.R. Russell}
\affiliation{School of Physics \& Astronomy, University of Nottingham, University Park, Nottingham NG7 2RD, UK}

\author{Prathamesh D. Tamhaneh}
\affiliation{Department of Physics and Astronomy, The University of Alabama in Huntsville, 301 Sparkman Dr NW, Huntsville, AL 35899, USA}

\author{S. Peng Oh}
\affiliation{Department of Physics, University of California, Santa Barbara, CA 93106, USA}

\author{Serena Perrotta}
\affiliation{Department of Astronomy and Astrophysics, University of California, San Diego, La Jolla, CA 92093, USA}




\begin{abstract}
We investigate the velocity structure of nebular gas in the central galaxies of four clusters: Abell 1835, PKS 0745-191, Abell 262, and RXJ0820.9+0752, using data from the Keck Cosmic Web Imager (KCWI). Velocity structure functions (VSFs) of the [OII] emission line are compared to VSFs of molecular clouds observed with the Atacama Large Millimeter/submillimeter Array (ALMA). Apart from Abell 262 where the gas is located in a circumnuclear disk, the nebular gas in the remaining galaxies lies in off-nuclear filamentary structures with VSFs steeper than the Kolmogorov slope. This steepening may be plausibly attributed to gravity although other factors, {such as magnetic stresses and bulk motion,} may be significant. The VSFs of CO and [OII] emission are similar in RXJ0820 and Abell 262, indicating close coupling of the nebular and molecular gases. In contrast, the nebular and molecular gases are differentiated on most scales in PKS 0745 and Abell 1835. This discrepancy is likely due to the radio-AGN churning the gas. We compare the scale-dependent velocity amplitudes of the hot atmospheres constrained by X-ray surface brightness fluctuation analysis using Chandra observations to the nebular VSFs. The large-scale consistency in Abell 1835 and RXJ0820 is consistent with condensation from the hot atmospheres. {We explore substantial systematic biases,} including projection effects, windowing, and smoothing effects when comparing VSFs using different telescopes and instruments.
\end{abstract}


\section{Introduction}

Galaxy clusters, the universe's largest gravitationally bound structures, are filled with ionized hot atmospheres with temperatures between from $10^7$ K and $10^8$ K. Within the cores of these clusters, an increase in gas density is expected to lead to substantial accumulations of cold gas, enable the condensation of molecular clouds at rates exceeding $\rm 100~M_\odot~{yr}^{-1}$, and thereby foster vigorous star formation but is unseen \citep[]{1977ApJ...215..723C,cavagnolo2008entropy,rafferty2008regulation,peterson2006x}. Despite central radiative cooling timescales often being shorter than the ages of clusters, their hot atmospheres are globally stable, remaining in both hydrostatic and thermal equilibrium \citep[]{mccourt2012thermal}. One or more heat sources are required to maintain this stability. The most prevalent heating mechanism is radio-mechanical Active Galactic Nucleus (AGN) feedback, in which the energy is injected from central supermassive black holes (SMBHs), interacting with the intercluster medium (ICM) via jet-inflated bubbles of relativistic plasma, seen as cavities in X-ray images \citep[]{voit2005observationally,2012ARA&A..50..455F,birzan2012duty}. The AGN energy associated with the cavity alone is enough to offset radiative cooling, by measuring the
surrounding pressure and volume of the cavities using the X-ray
data \citep[]{rafferty2006feedback}. However, the process by which this energy is transferred to the ICM  is not {well understood.}

Turbulence is thought to be an important channel for coupling feedback to the environment \citep[]{churazov2002cooling,omma2004heating,gaspari2014relation}. The analysis of X-ray surface brightness (SB) fluctuations indicates that the ICM in the cluster cores is turbulent, consistent with velocities consistent with the Hitomi Doppler line broadening measurements \cite[]{hitomi2016quiescent}. Turbulent dissipation {may be sufficient to compensate for cooling losses, although measurement uncertainties prevent a conclusive determination} \citep[]{zhuravleva2014turbulent}. Other possibilities for channeling energy to the ICM include shocks \citep[]{heinz1998x,fabian2006very,2007PhR...443....1M,randall2010shocks}, sound waves \citep[]{sternberg2009sound,sanders2007deeper,2013AN....334..386N}, streaming and diffusion of cosmic ray protons \citep[]{1991ApJ...377..392L,guo2008feedback,pfrommer2013toward}, radiative heating \citep[]{nulsen2000fuelling,ciotti2001cooling}, and mixing of gas between the ICM and the hot bubbles \citep[]{hillel2016heating,hillel2017hitomi}. {Many or all may contribute to heating.} 

Despite suppression of cooling by heating processes in groups and clusters, the detection of cool-phase gas within the atmospheres of the brightest cluster galaxies (BCGs) and giant elliptical galaxies indicates significant cooling.  Massive molecular gas reservoirs over $\rm 10^9~M_\odot$ are observed near the nucleus of some clusters, which fuels star formation and sustains the feedback loop \citep[]{edge2001detection,salome2003cold,rafferty2008regulation,russell2019driving,olivares2019ubiquitous}. ALMA observations of central galaxies in clusters show that the molecular gas preferentially lies around and in the wakes of buoyantly rising bubbles \citep[]{mcnamara20141010,russell2017close,russell2019driving,olivares2019ubiquitous}. However, if and how the molecular gas couples to the bubbles is not understood. The cold dense gas may be accelerated outward from the center. Alternatively, with the difficulty of lifting high column density clouds, the molecular gas might cool from the hotter gas lifted behind bubbles \cite[]{mcnamara2016mechanism}. Moreover, optical line emission at $\sim 10^4$ K is commonly observed in the centers of cool-core clusters \citep[]{1989ApJ...338...48H,crawford1999rosat,cavagnolo2008entropy,mcdonald2011effect,gingras2024complex}. This filamentary nebular emission, tracing warm ionized envelopes of many cold molecular gas clouds \citep[]{jaffe2005h}, wraps around both the radio jet and the X-ray cavities \citep[]{salome2008cold,mcnamara20141010,vantyghem2016molecular,russell2017close}. Such a spatial coupling is consistent with a top-down multi-phase condensation cascade, within which both the warm ionized and cold molecular components are cooling products that rain from the hot ambient plasma. Their kinematics are predicted to retain the ‘memory’ of the hot gas from which they condense \citep[]{gaspari2017raining,gaspari2018shaken,voit2018role}.

The velocity structure function (VSF) provides a powerful tool to characterize gas motions and dynamical processes. The kinematics of multi-phase filaments in three nearby galaxy clusters,  Perseus, Abell 2597, and Virgo, have been examined by \cite{2020ApJ...889L...1L}. Their results indicate that the cold filaments in these clusters are well coupled to the turbulent nature of the hot gas that is present. The agreement between the turnover scales of the VSF and the dimensions of the observed bubbles inflated by jets suggests that activity from the central supermassive black hole drives turbulent gas motions in cluster cores, highlighting the potential role of turbulence in channeling feedback energy to the ICM. In contrast, \cite{hillel2020kinematics} noticed that the VSFs of cold filaments in these three clusters were steeper than the Kolmogorov slope of 1/3, typical for classical turbulent cascades, leading them to claim that turbulence may not be a major heating mechanism in these clusters due to long dissipation timescales. Instead, they concluded that mixing with hot gas from the bubbles is more effective in counteracting radiative cooling. By comparing with 3D hydrodynamical simulations of jet-inflated hot bubbles, as studied in \citep[]{hillel2016heating}, they suggested that jets could directly induce turbulence with a VSF slope steeper than 1/3. By examining magnetohydrodynamical (MHD) simulations of self-regulated AGN feedback in a Perseus-like cluster, \cite{wang2021non} found that the cold phase (T $< 10^4$ K) VSF was steeper than predicted by Kolmogorov’s theory and attributed this steepening to gravitational acceleration acting on cold clouds. The results imply that the turbulence in the surrounding hot medium may be driven by the motion of precipitating cold filaments and by AGN jets. Furthermore, by varying initial outflow properties in hydrodynamical simulations, \cite{2022ApJ...929L..30H} discovered that a supersonic turbulent velocity structure with VSF slope of 1/2, can be generated and be 'frozen' within the $\rm H\alpha$-emitting filaments emerging from fast AGN-driven hot outflows. Gravitational interaction tends to flatten the VSF of the cold phase over a short timescale, approximately 10 Myr, indicating that the lack of flattened VSF reflects the short-lived nature of the $\rm H\alpha$ emitting phase.

{The nebular emission from the core regions of four galaxy clusters—Abell 1835, Abell 262, PKS 0745-191, and RXJ0820.9+0752—has been analyzed using new optical observations from the Keck Cosmic Web Imager (KCWI) \citep{morrissey2018keck, gingras2024complex}. As the [OII] emission line doublet is the highest luminosity line observed in the KCWI data,  \cite{gingras2024complex} focused on the  [OII]3726,9 \AA{}, also incorporated data from previous multiwavelength studies, including radio and X-ray observations, to examine the morphology and dynamics of the nebular gas. In this work, we continued our VSF analysis by utilizing this integral field spectroscopy (IFS) data, which offers a larger field-of-view (FOV) and relatively high spectral resolution compared to earlier studies. This enables us to effectively probe the dynamical processes affecting the gas across a broad range of spatial scales. Furthermore, the velocity structures of the multi-phase filaments may reflect the origins of the filaments and the interactions between AGN feedback and the intracluster medium (ICM). Therefore, we have also incorporated multi-wavelength observations to provide a detailed analysis of the kinematics of both molecular gas and the hot gas, and to conduct a thorough comparison with the warm ionized gas.} 
The molecular gas in these clusters was {observed by ALMA} and is traced by J = 1 - 0, J = 2 - 1, and J = 3 - 2 rotational transitions of CO with high spatial and velocity resolution. Additionally, by comparing the VSF feature scales with the structures identified in the X-ray images observed by the Chandra X-ray Observatory, we can explore the correlations between their dynamics and the jet-inflated bubbles within these systems.

The line-of-sight velocity dispersion of the hot ICM has been measured directly in the core of Perseus by Hitomi \citep{hitomi2016quiescent} with relatively low spatial resolution. Under certain assumptions, the scale-dependent velocity amplitude of the hot gas motion can be derived indirectly from the measured power spectra of X-ray SB fluctuations. In this work, we analyze the velocity power spectra of hot gas motion obtained from Chandra observations and compare them to the VSFs of filaments observed with KCWI. This analysis provides us with insights into the role of turbulence in reheating the ICM and the connection between the motion of gas in different phases and the activity of the central SMBHs.

In Section~\ref{sec:data} we introduce the central galaxies studied in this work and provide a summary of the data processing procedures. In Section \ref{sec:result} we present the results of the VSF measurements based on KCWI and ALMA observations, analyzing their shapes, slopes, and amplitudes, and comparing them to Kolmogorov turbulence theory and hot ICM motions obtained from X-ray SB analysis. In Section \ref{sec:discussion} we discuss the shape of measured VSFs, as well as the the interaction between gas in different phases. Finally, we conclude this work in Section \ref{sec:conclusion}. Throughout the paper we assume a standard $\rm \Lambda CDM$ cosmology with $H_0= 70~\rm km~s^{-1}~Mpc^{-1}$, $\Omega_{\rm m}=0.3$, and $\Omega_{\rm \Lambda}=0.7$.

\section{Cluster Sample and Observational Data}

\label{sec:data}

\subsection{Cool Cluster Sample}

{Though limited to only four clusters, the sample in this study is adequately representative of the broader cool cluster population and was chosen to encompass a broad range of feedback properties.} The details are listed in the Table~\ref{tab: samples}. The redshifts lie between $z=$ \(0.0165\) and $z=$ \(0.2512\). The cluster atmospheres have central cooling times below \(1 \text{Gyr}\) and harbor substantial reservoirs of molecular gas with masses ranging between $\rm 10^8~M_\odot$ and $\rm 10^{10}~M_\odot$. The central galaxy of Abell 1835 is experiencing powerful radio-mode feedback and has a high star formation rate \citep[]{mcnamara2006starburst}. A powerful radio source is in the core of PKS 0745, more than ten times higher than those of other clusters in our sample. CO emission emitted from molecular gas in these clusters spatially overlaps the nebular gas in all sample clusters, allowing us to directly compare the kinematics of these gas phases. In RXJ0820, the molecular and nebular gases show a significant spatial offset from the central galaxy. Both gas phases in Abell 262 exhibit a disk-like structure that rotates around the nucleus. Combined with the detection of X-ray cavities in these clusters, which result from previous central AGN activity, the multi-wavelength observations allow us to study the dynamics of different phases of the ICM and their interconnection in the AGN feedback model.

\subsection{KCWI Data Reduction}
\label{sec:keck_data_reduction}

Details of the KCWI observations and data reduction of these four central galaxies are presented in \cite{gingras2024complex}; here we report the most relevant details. The properties of KCWI observations are given in column (2) - (4) of Table ~\ref{tab:observation}. 
The data was reduced using the KCWI IDL Data Extraction and Reduction Pipeline \texttt(KDERP). The data was resampled onto a \(0.29'' \times 0.29''\) spaxel grid, and a mosaic was created from the individual pointings observed for each cluster.

The stellar continuum and emission lines in each spaxel were simultaneously fit using the \texttt{IFSFIT} IDL library. To obtain the stellar velocity in the innermost 1" region of the central galaxy, which was used as the systemic velocity of the system, a spatially-integrated spectrum was created by summing the spectra from spaxels with a strong stellar continuum in the central galaxy. Emission lines were masked out and the stellar continuum was fit using the Penalized Pixel-Fitting method \texttt{PPXF} \citep{PPXF}. The stellar component is matched to stellar population synthesis (SPS) models \citep{SSP2005}. The best-fit SPS models across all stellar ages were then summed to obtain the stellar continuum fit in the spatially-integrated spectrum.

{The} stellar continuum and emission lines were fit simultaneously. Using \texttt{MPFIT} \citep{MPFIT}, the line profiles were convolved with the KCWI spectral resolution to fit the emission lines. {The flux ratio of [OII] doublet was fixed at 1.2, based on an assumed electron density of \(400 \, \text{cm}^{-3}\) \citep{mclaughlin1998electron,pradhan2006oii}.} A flux threshold of signal-to-noise ratio (S/N) $>$ 5 per spaxel was applied to all the flux and kinematic maps of the [OII] 3726,9 \AA{} doublet presented here.

In performing the emission line fits, a second component is included when the Bayesian Information Criterion (BIC) for the two-component fit is less than that for the one-component fit and when neighboring spaxels also require a second kinematic component. Of the four cooling clusters observed, only one kinematic component is required for Abell 262 and RXJ0820, while a second kinematic component is needed in Abell 1835 and PKS 0745. More details are provided in \cite{gingras2024complex}. 

\begin{figure*}
\centering
	\includegraphics[width=15cm,height=14cm]{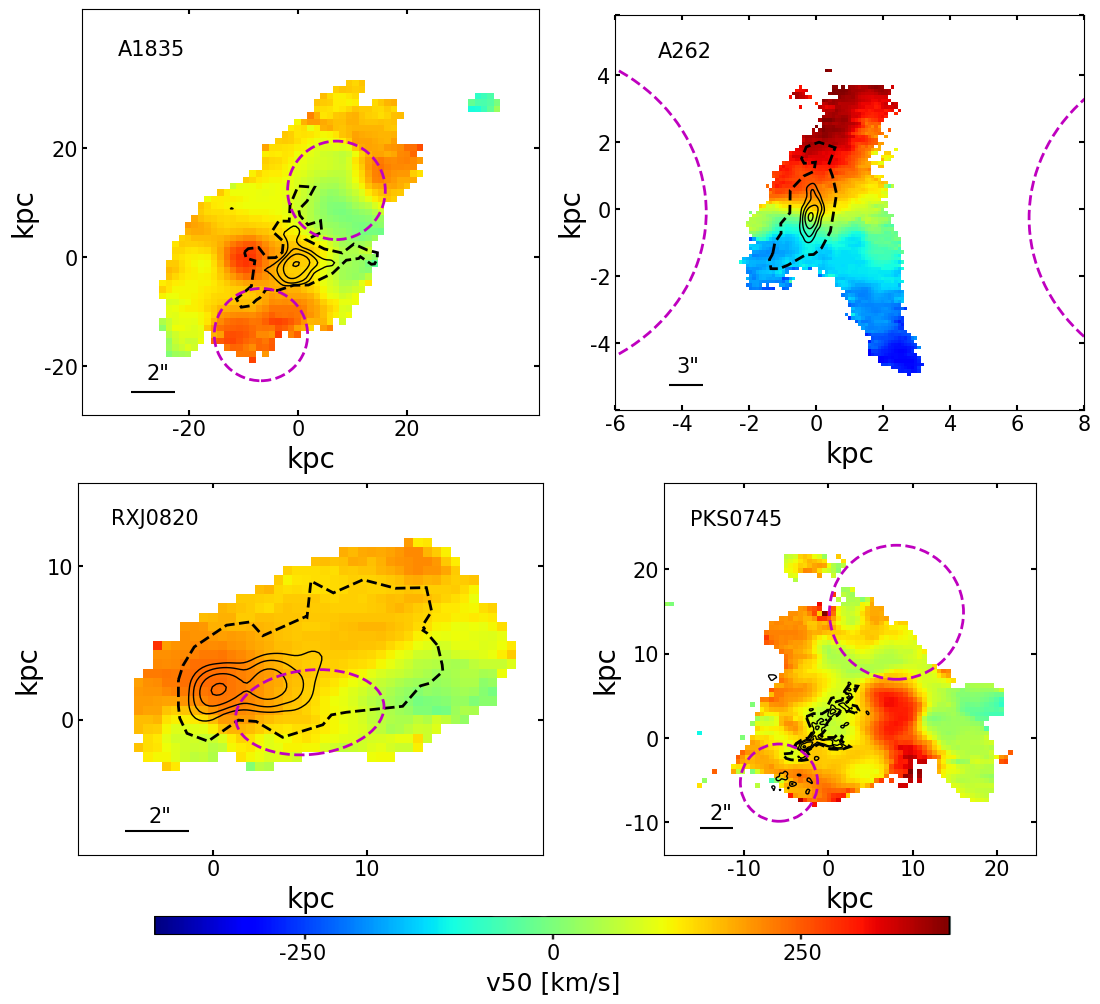}

    \caption{The median velocity v$_{50}$ maps of the [OII] emission line doublet with respect to the stellar velocity of the system. The central galaxy nucleus is designated as the reference point for each map's origin. Positive velocities indicate redshift, while negative velocities denote blueshift relative to the stellar velocity. The black solid lines indicate the integrated flux contours for 10\%, 20\%, 40\%, and 80\% of the maximum CO (3-2) emission in Abell 1835 and PKS 0745, CO (2-1) emission in Abell 262, and CO (1-0) emission in RXJ0820. Areas within the black dashed lines highlight regions where CO emissions are detected by ALMA. The dashed magenta ellipses show the positions of the X-ray cavities.}
    \label{fig: keck_v50_maps}
\end{figure*}

{Figure \ref{fig: keck_v50_maps} presents the median velocity v$_{50}$, which is the 50th percentile of the cumulative velocity profiles of the [OII] emission line doublet in the four cooling cluster cores.} Positive velocities indicate a redshift when compared to the systemic redshift of the central galaxy, while negative velocities are blueshifted. {The nucleus of the central galaxy, located at the origin of the images, corresponds to the brightest spaxel in the stellar continuum flux map of Abell 1835, PKS 0745 and RXJ0820. The location of the nucleus in Abell 262 is determined from the spaxel with the highest stellar velocity dispersion, as a strong central dust lane leads to substantial extinction in this central galaxy.} Most of the nebular gas is redshifted with respect to the stars by approximately $150~\rm km~s^{-1}$ in Abell 1835, $135~\rm km~s^{-1}$ in PKS 0745, and $150~\rm km~s^{-1}$ in RXJ0820. The nebular gas in Abell 262 is blueshifted in the south and redshifted in the north, with median velocities ranging between $-~314~\rm km~s^{-1}$ and $+~422~\rm km~s^{-1}$. This clear velocity gradient demonstrates ordered motion along the north-south axis.

\begin{deluxetable*}{lccccccc}
    \tablecaption{Target details \label{tab: samples}}
    \tabletypesize{\scriptsize}
    \tablewidth{\textwidth} 
    \tablehead{
        \colhead{Target} & \colhead{R.A.} & \colhead{Dec.} & \colhead{z} & \colhead{a} & \colhead{b} & \colhead{d} & \colhead{$\rm M_{mol}$} \\
        \colhead{} & \colhead{(J2000)} & \colhead{(J2000)} & \colhead{} & \colhead{(kpc)} & \colhead{(kpc)} & \colhead{(kpc)} & \colhead{($\rm M_{\odot}$)}
    }
    \colnumbers
    \startdata
    Abell 1835 & 14:01:02.1 & +02:52:43 & 0.2514 & 14 & 10 & 17  & 5 $\times 10^{10}$ \\
               &            &           &        & 16 & 12 & 23 & \\
    Abell 262  & 01:52:46.5 & +36:09:07 & 0.0160 & 11 & 10 & 8 & 3.4 $\times 10^8$ \\
               &            &           &        & 11 & 10 & 10 & \\
    RXJ0820.9+0752 & 08:21:02.3 & +07:51:47 & 0.1103 & 9 & 5 & 6 & 3.9 $\times 10^{10}$ \\
    PKS 0745-191 & 07:47:31.3 & -19:17:40 & 0.1024 & 10 & 10 & 6 & 4.6 $\times 10^9$ \\
                 &            &           &        & 17 & 17 & 20 & \\
    \enddata
    \tablecomments{Targets details: (1) Cluster name. (2) Right ascension (RA). (3) Declination (Dec). (4) Redshift (5) Major axis of X-ray cavity. (6) Minor axis of X-ray cavity. (7) Distance from cavity center to the core. (8) Molecular gas mass. References: (5)-(7): Abell 1835: \cite{mcnamara2006starburst, olivares2019ubiquitous}, Abell 262: \cite{rafferty2006feedback,clarke2009tracing}, RXJ0820.9+0752: \cite{vantyghem2019enormous}, PKS 0745-191: \cite{sanders2014feedback}.}
\end{deluxetable*}

\begin{deluxetable*}{lcccccccc}
    \tablecaption{The observations used in this analysis }
    \label{tab:observation}
    \tablewidth{\textwidth}
    \tablehead{
        \colhead{Target} & \multicolumn{3}{c}{KCWI Observations} & \multicolumn{3}{c}{ALMA Observations} & \multicolumn{2}{c}{Chandra Observations} \\
        \colhead{} & \colhead{Seeing} & \colhead{FOV} & \colhead{Emission Extent} & \colhead{CO Line} & \colhead{Beam Size} & \colhead{Emission Extent} & \colhead{ObsIDs} & \colhead{Clean $\rm T_{exp}$} \\
        \colhead{} & \colhead{(")} & \colhead{(" $\times$ ")} & \colhead{(kpc $\times$ kpc)} & \colhead{} & \colhead{(" $\times$ ")} & \colhead{(kpc $\times$ kpc)} & \colhead{} & \colhead{(ks)}
    }
    \decimalcolnumbers
    \startdata
    Abell 1835  & 0.8 & 35 $\times$ 33 & 62 $\times$ 51 & J = 3 - 2 & 0.45 $\times$ 0.6 & 25 $\times$ 22 & 6880, 6881, 7370 & 178 \\
    Abell 262   & 0.9 & 30 $\times$ 32 & 5.5 $\times$ 8.9 & J = 2 - 1 & 0.6 $\times$ 1.0 & 2 $\times$ 4 & / & / \\
    RXJ0820.9+0752 & 0.9 & 21 $\times$ 24 & 41 $\times$ 26 & J = 1 - 0 & 0.7 $\times$ 0.7 & 18 $\times$ 11 & 17194, 17563 & 60 \\
    PKS 0745-191 & 1.2 & 26 $\times$ 31 & 33 $\times$ 30 & J = 3 - 2 & 0.2 $\times$ 0.3 & 11 $\times$ 14 & 2427, 6103, 7694, & 176 \\
    & & & & & & & 12881, 19572 - 5 & \\
    \enddata
    \tablecomments{Columns (1) Cluster name. {KCWI observations}: (2) Seeing. (3) {Total field of view of multiple pointings.} (4) [OII] emission extent. {ALMA observations}: (5) CO emission line used for fitting. (6) The synthesized beam size. (7) CO emission extent. {Chandra observations}: (8) Chandra ObsIDs used in this work. (9) Cleaned exposure time.}
\end{deluxetable*}

\subsection{ALMA Observations}

We use archival ALMA observations of the four cooling cluster cores in sample to compare the dynamic properties of the cold molecular gas with the warm ionized gas. Details of these observations are listed in column (5) - (7) of Table~\ref{tab:observation}. The data were calibrated using the ALMA pipeline reduction scripts with \texttt{CASA} version 4.7.2. Standard phase calibration was performed, and the data cube was binned in $\rm 10~km~s^{-1}$ resolution. 
{The black dashed contours in Figure~\ref{fig: keck_v50_maps} represent the region of the CO emission detected at a 3$\sigma$ level by ALMA.} While the observed molecular gas spatially overlaps with the nebular gas, it is more concentrated near the central galaxy of each cluster \citep{gingras2024complex}. 

\begin{figure*}
\centering
	\includegraphics[width=18cm,height=17cm]{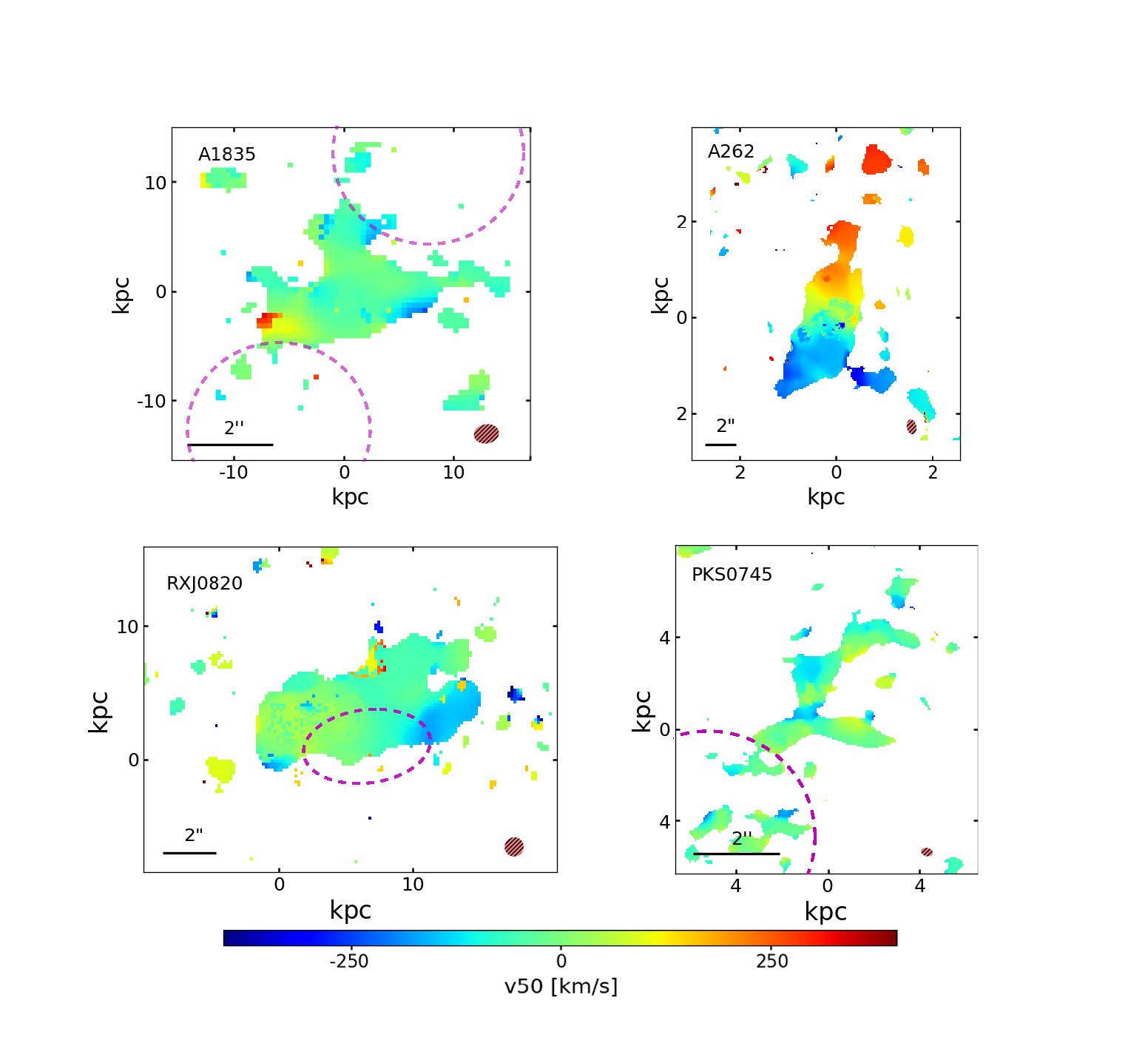}

    \caption{Median {velocity} maps of molecular gas in the central regions of four galaxy clusters from our sample, derived from ALMA observations. The maps show CO (3-2) emission for Abell 1835 and PKS 0745, CO (2-1) emission for Abell 262, and CO (1-0) emission for RXJ0820, all detected at a signal-to-noise ratio greater than $\rm 3\sigma$. {Map origins align with the nucleus positions defined in the KCWI [OII] maps (presented in Figure \ref{fig: keck_v50_maps}).} Red hatched ellipses represent the synthesized beam size and dashed magenta ellipses indicate the locations of X-ray cavities.}
    \label{fig: alma_v_maps}
\end{figure*}

Figure~\ref{fig: alma_v_maps} presents the velocity maps of the molecular gas traced by CO emission. {These velocity fitting results are published in \cite{russell2019driving}.} Gaussian smoothing over the synthesized {beam was} applied. The origins are consistent with those in Figure~\ref{fig: keck_v50_maps}, indicating the locations of the nuclei of the central galaxies. CO emission shown was detected at a $\rm S/N > 3\sigma$ threshold. When multiple CO lines are detected, we present either the stronger line or the line with higher spatial resolution over a larger area, whichever is more suitable for our purposes. The specific CO emission line used for each object is listed in column (5) of Table \ref{tab:observation}.

\subsection{Chandra X-ray Observations}
\label{sec:chandra_obsid}

To investigate the dynamics of the hot atmospheres in the cores of these clusters, we measured its velocity power spectra within the KCWI [OII] emission regions based on archival Chandra X-ray data available in the archive, observed by the Advanced CCD Imaging Spectrometer (ACIS). Details of the Chandra observations are listed in column (8) - (9) of Table~\ref{tab:observation}. Each observation was reduced by \texttt{CIAO} 4.15 with the latest \texttt{CALDB 4.15} calibration, following the standard algorithm described by \cite{vikhlinin2005chandra}. Intervals contaminated by background flares were removed from each observation using the light curves extracted from \texttt{level-2} event files above 10 keV. The X-ray background files were estimated using blank sky files, which were re-normalized to align with the count rates in the 10-12 keV band. To measure the gas density fluctuations, the exposure-corrected, background-subtracted images of each observation were generated in the $0.5 - 3.5~\rm keV$ band. In this energy range, the X-ray emissivity is independent of gas temperature for clusters with a mean temperature T $>\sim$3 keV. For each target, all observations were reprojected onto the observation with the largest exposure time and summed up into a single mosaic image. The bright point sources identified using \texttt{wavdetect} tool, as well as the central AGNs, were masked out from subsequent SB analysis.


\section{Results}
\label{sec:result}

{In Section \ref{sec:3.1}, we first introduce the methods used for VSF analysis in this study. Section \ref{sec:3.2} presents the VSFs of $v_{50}$ for four central galaxies. Each Gaussian velocity component is analyzed for Abell 1835 and  PKS 0745 in Section \ref{sec:components}. Comparisons between the VSFs of warm ionized gas and the cold molecular gas traced by ALMA CO emission are detailed in Section \ref{sec: alma vsf}. Additionally, velocity spectra of hot gas observed by Chandra X-ray observations are discussed in Section \ref{sec:3.5}.}

\subsection{Velocity Structure Function}
\label{sec:3.1}
The VSF is a powerful analytical tool that extends beyond turbulence studies. It provides insights into the kinematics of gas, allowing us to identify and quantify the physical processes driving the observed motions such as rotational dynamics, gravitational effects, and interactions with external forces.

For VSF calculation, the projected separation, denoted by \( l \), and the absolute velocity difference \( |\delta v| = |v_i - v_j| \) is determined for all possible combinations of selected pixels. These pixel pairs are then grouped into distinct bins based on their separation \( l \). Within each bin, the average of the velocity differences \( \langle |\delta v| \rangle \)is computed to determine the VSF amplitude at that specific scale. We experimented with evenly grouping pairs across the scales of interest and confirmed that the VSF results remain consistent. 

The measurements are taken only at scales exceeding the point spread function (PSF). The bins with an insufficient number of pixel pairs can undermine the reliability of VSF measurements, especially at larger separations where the constraints of the image size become a limiting factor. To address this, bins where pairs drop below 20\% of the peak value are excluded. {This threshold ensures that sufficient pairs are included at each separation.} {The turnover scale refers to the point where the VSF amplitude reaches a local maximum. This scale may indicate the scale where energy is injected into the system and subsequently dissipates onto smaller scales, or it may be related to other factors such as bulk motions.} Our findings indicate that the VSF is reliable up to scales no larger than half the extent of the map. {Significant slope changes or turnovers may be influenced by insufficient pixel pairs at large separations — a limitation resulting from the restricted image size. This phenomenon, referred to as the `window effect', stems from statistical uncertainty due to limited data size.}
 The impact of the window effect on VSF measurements is further explored and discussed in Appendix~\ref{appdenix:window}.

The velocity differences \(|\delta v|\) spread over a wide range within each bin. {Simply using the mean difference does not adequately represent the \(|\delta v|\) among all pairs.} Therefore, to obtain a more comprehensive understanding, we analyze the \(|\delta v|\) distribution at each measured scale. This detailed examination enhances our understanding of the characteristics of the resulting VSFs as well as the kinematics of gas within the cluster cores. We group the pairs into $10~ \rm km~s^{-1}$ bins, ranging between 0 and the maximum \(|\delta v|\) for each object. To allow for an efficient comparison analysis, each \(|\delta v|\) bin count is normalized by the total number of pairs at that scale, ensuring value ranges between 0 and unity, allowing for comparison across scales.

Three of the clusters exhibit distributions with a peak at lower values and an extended tail reaching toward higher values. This shape resembles a log-normal distribution, which describes a dataset where the logarithm of the values follows a Gaussian distribution. The \(|\delta v|\) distribution in each separation bin was fitted by a log-normal profile. Using the parameters from the best-fit model, the VSFs we reconstructed based on the mean of the log-normal distribution. However,  the $|\delta v|$ distribution of the disk in Abell 262 does not fit a log-normal distribution; more details will be discussed in Section~\ref{sec:3.2}. To maintain consistency across objects and to allow comparisons with previous studies, we continue to use the averaging method for calculating the VSF. Nevertheless, modeling the $|\delta v|$ distributions and not just the VSF alone may provide better constraints and allow for more detailed comparisons to simulations than VSFs alone.


\subsection{VSF of warm ionized gas}
\label{sec:3.2}
\subsubsection{Abell 1835}
\label{sec:Abell 1835}

Abell 1835 is a typical example of a cool core cluster where the central galaxy is undergoing intense radio-mode feedback. X-ray observations have revealed two cavities as surface brightness depressions: one with a radius of $\sim$16 kpc, located $\sim$23 kpc northwest of the center, and another with a radius of 14 kpc, situated $\sim$17 kpc to the southwest \citep{mcnamara2006starburst}. Furthermore, Abell 1835 is characterized by strong nebular emission, with an H\(\alpha\) luminosity of $\rm 1.7\times 10^{42}$ erg/s \citep{wilman2006integral}. The [OII] emission traced by KCWI extends over a wide area, reaching beyond $\sim 45$ kpc in right ascension and $\sim$ 50 kpc in declination. As seen in the upper left panel of Figure \ref{fig: keck_v50_maps}, the nebular gas closely reflects the alignment of the X-ray bubbles, extending approximately 50 kpc along the axis of the bubbles. The median velocity of the [OII] emitting gas ranges between -30 $\rm km~s^{-1}$ and +290 $\rm km~s^{-1}$. Most nebular gas is redshifted, indicating a consistent movement away from the observer relative to the central stellar velocity \citep{gingras2024complex}.

\begin{figure*}
\centering
	\includegraphics[width=\textwidth]{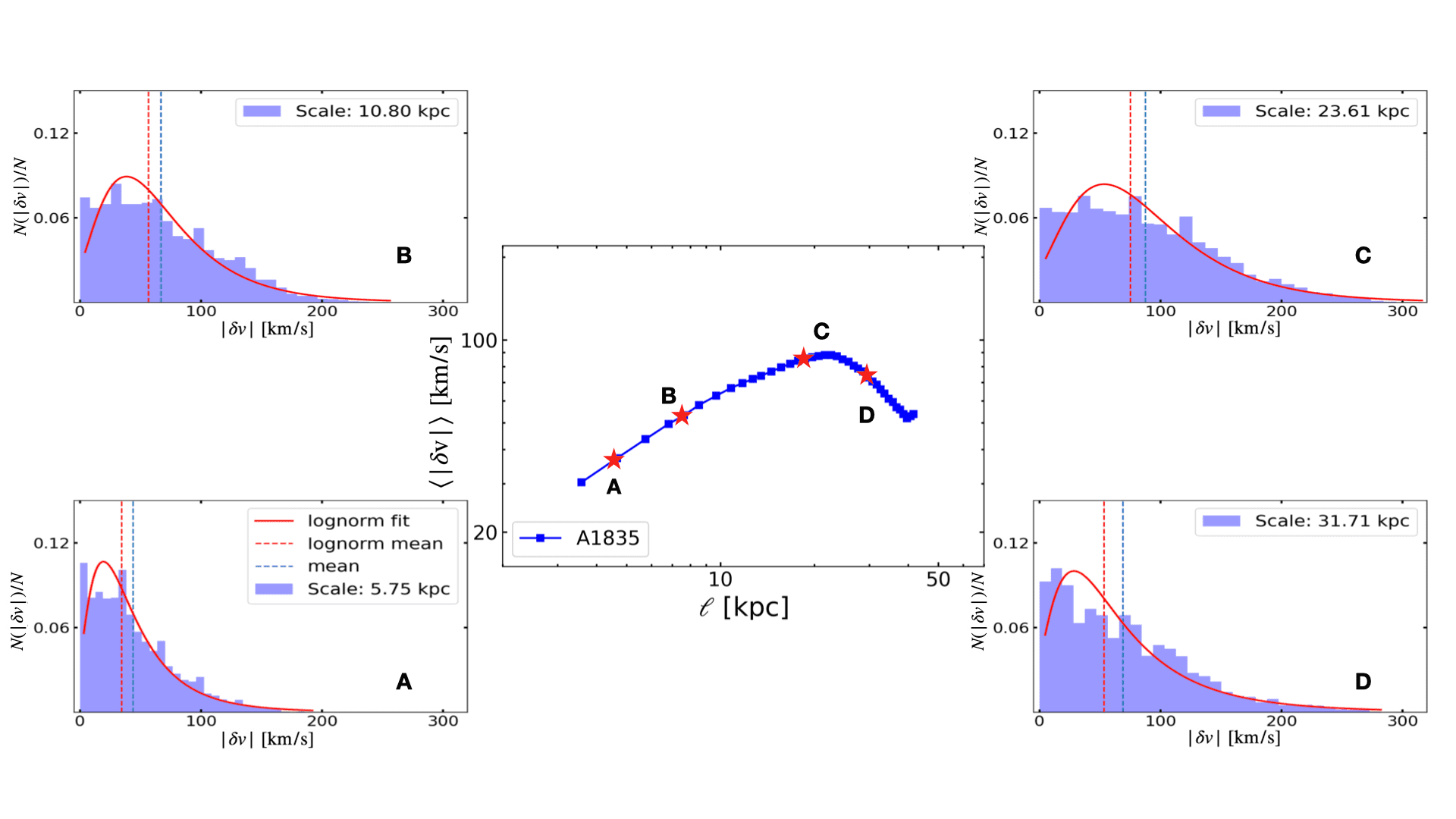}
    \caption{\textbf{Middle:} VSF analysis of warm ionized gas, traced by KCWI [OII] emission, within the core of Abell 1835. {Only scales larger than seeing, and where pairs exceed 20\% of the peak value, are plotted.} \textbf{Side panels:} Distribution of velocity differences \(|\delta v|\) among pixel pairs at four scales, marked as red stars in the VSF plot. The number of pairs in each \(|\delta v|\) group is normalized by the total number of pairs for the entire separation bin to facilitate a comparative analysis of distribution shapes across varying scales. The best-fit log-normal profile is displayed as a solid red line over the \(|\delta v|\) distribution. The vertical red dashed line represents the mean of the log-normal distribution, while the {blue} line shows the mean \(|\delta v|\) of all pixel pairs within the given bin.}
    \label{fig:keck_vfs_a1835}
\end{figure*}

The middle panel of Figure~\ref{fig:keck_vfs_a1835} displays the VSF of nebular gas in Abell 1835. The VSF ranges from 30 $\rm km~s^{-1}$ on the smallest scale of $\sim$4 kpc to 88 $\rm km~s^{-1}$ at a scale of 22 kpc. {The turnover presented at a scale of 22 kpc suggests the driving scale of turbulence, which corresponds to the distance from the X-ray bubble to the nucleus of Abell 1835, indicating a potential link to radio-mechanical AGN feedback.} Below this scale, the VSF increases steadily with a slope of approximately 0.39 at scales between 10 and 20 kpc and steepens at scales below 10 kpc. A summary of the [OII] VSF measurements for Abell 1835, as well as those for the other three clusters, is presented in Table~\ref{tab:keck_vsf}.

The side panels of Figure~\ref{fig:keck_vfs_a1835} illustrate the distributions of velocity differences $|\delta v|$ at four distinct scales to present the characteristic shapes as the scale varies. The mean $|\delta v|$ of all pairs is indicated by the blue vertical dashed line, corresponding to the VSF amplitude, marked by a red star in the middle panel. The histograms typically show a left-skewed peak towards the lower velocity differences, while the extended tails indicate a significant number of pixel pairs with high $|\delta v|$. As the separation scale grows from panels A to D, the tail of the distribution broadens, with $|\delta v|$ values reaching up to approximately $\rm 290 ~km~s^{-1}$ at a scale of 32 kpc. This trend is consistent with the expectation that larger separations correlate with higher velocity differences. However, at smaller scales, such as the 5.75 kpc scale shown in panel A, a notably high $|\delta v|$ tail extends to nearly $\rm 200 ~km~s^{-1}$, far exceeding the mean of $\rm \sim 44 ~km~s^{-1}$. This may be attributed to projection effects where pixel pairs close in projection but not in three-dimensional space exhibit large velocity differences. The red solid line in the histograms represents a log-normal fit to the distribution of $|\delta v|$. The means derived from the log-normal fits, denoted by the vertical red dashed lines, are consistently $\sim$14\% to 22\% lower than the direct mean of all pairs, suggesting that VSFs reconstructed by log-normal fitting can mitigate the influence of the extended tails, especially at smaller scales where these tails are more pronounced. Based on the comparison of the best-fit profiles and the distributions, the log-normal fit provides a reasonable approximation of these distributions in Abell 1835.

\begin{deluxetable}{lCCCC}
\tablecaption{VSF of KCWI [OII] emission line\label{tab:keck_vsf}}

\tablewidth{\textwidth}
\tabletypesize{\footnotesize}
\tablehead{
    \colhead{Target} & 
    \colhead{Scales} & 
    \colhead{Turnover scale} & 
    \colhead{Scales for fit} & 
    \colhead{Slope} \\
    \nocolhead{} & 
    \colhead{(kpc)} & 
    \colhead{(kpc)} & 
    \colhead{(kpc)} & 
    \nocolhead{} \\
    \colhead{(1)} & 
    \colhead{(2)} & 
    \colhead{(3)} & 
    \colhead{(4)} & 
    \colhead{(5)}
}
\startdata
Abell 1835 & 3.6 - 43 & 22 & 3.6 - 22 & 0.58 \\
           &          &    & 3.6 - 12 & 0.70 \\
           &          &    & 12 - 22  & 0.39 \\
Abell 262  & 0.6 - 5.9 & / & 0.6 - 5.9 & 1.0 \\
RXJ0820.9+0752 & 2 - 17 & 9 & 2.0 - 8.7 & 0.60 \\
PKS 0745-191 & 3.1 - 24 & 9 & 3.1 - 8.1 & 0.50 \\
\enddata
\tablecomments{Columns: (1) Target, showing the cluster name. (2)-(5) VSF Characteristics, including probed scales, turnover scale, scales for slope fitting, and best-fitting slope of the VSF.}
\end{deluxetable}

\subsubsection{Abell 262}

\begin{figure*}
\centering
	\includegraphics[width=\textwidth]{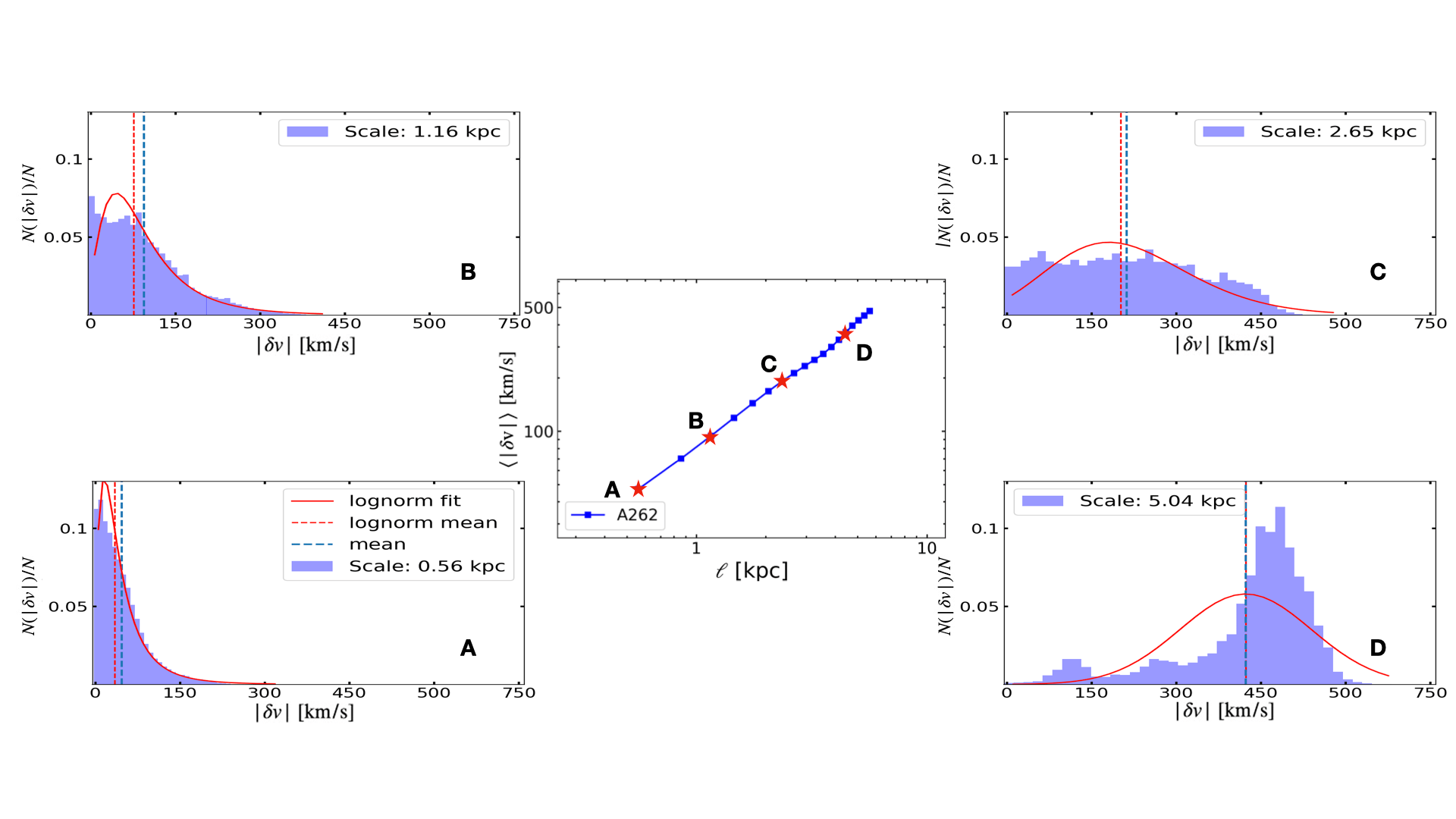}
    \caption{\textbf{Middle:} Similar to Figure~\ref{fig:keck_vfs_a1835}, the middle panel displays the VSF for the nebular gas within the core of Abell 262. {Side panels:} the distributions of $|\delta v|$ within four separation bins.}
    \label{fig:keck_vfs_a262}
\end{figure*}

VSF can be measured at scales below 1 kpc due to the low redshift of Abell 262 and are plotted and shown in the middle panel of Figure~\ref{fig:keck_vfs_a262}. The mean velocity differences, ranging between 47 $\rm km~s^{-1}$ at a 0.56 kpc scale and 510 $\rm km~s^{-1}$ at a scale of 5.9 kpc, are substantially higher than the other three targets which have velocity differences  $\sim$100 $\rm km~s^{-1}$. The continuous rise seen here with a slope of $\sim$1 suggests that the absolute velocity difference between two points increases linearly with their separation distance. {This pattern is typical for systems where the velocity structure is dominated by rotation, rather than random turbulent motion.}

The side panels of Figure~\ref{fig:keck_vfs_a262} display velocity difference distributions at four specific separations. At small scales (in panels A and B), the distributions fit well by a log-normal profile, shown as the red solid line. However, at intermediate scales (in panel C), the distribution begins to widen and flatten, suggesting a more diverse mixture of velocity differences due to the varied orientations of pixel pairs across the map. At larger scales (in panel D, $\sim$ 5 kpc), the peak of the distribution shifts toward higher \( |\delta v| \) values, marking a significant departure from the log-normal fit. This shift is indicative of increasing velocity differences with separation, consistent with the ordered, large-scale rotational motion within the ionized gas.

\begin{figure*}
\centering
	\includegraphics[width=17cm,height=6.5cm]{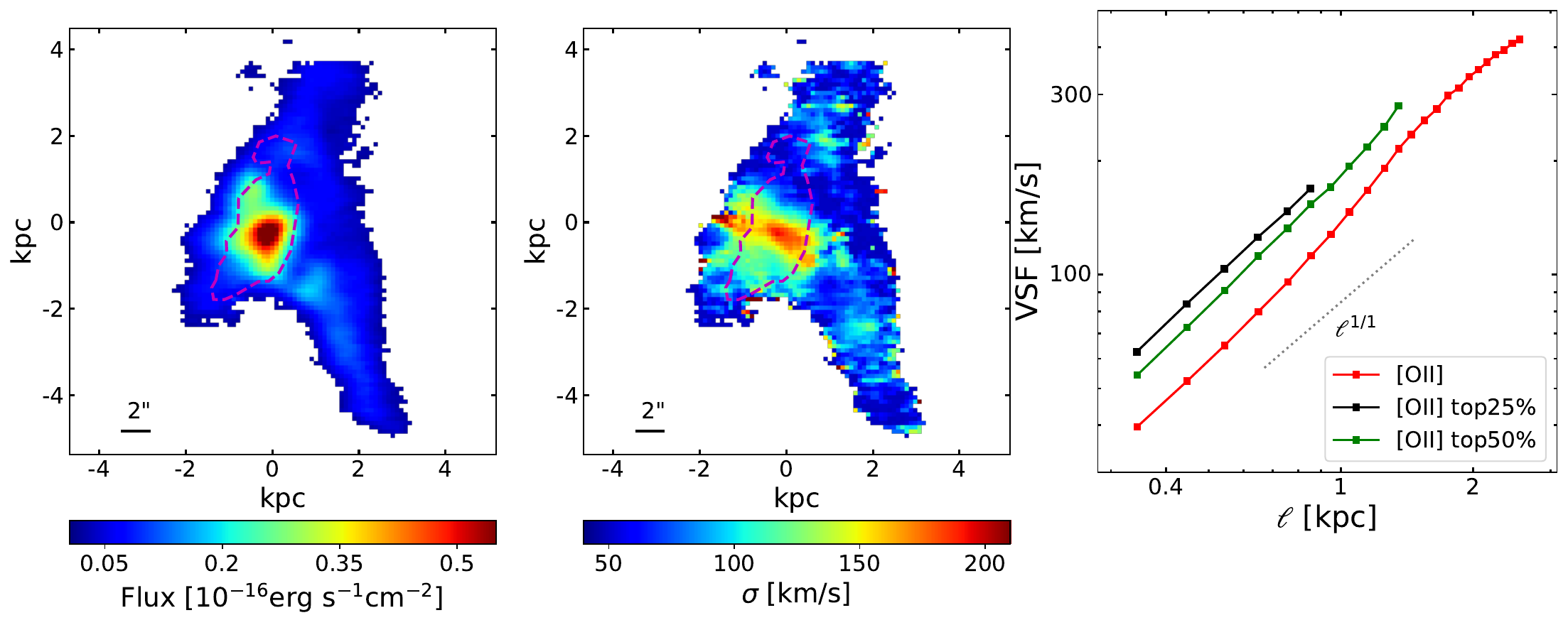}
    \caption{{Left panel:} Continuum-subtracted [OII] flux maps with $\rm S/N \ge 5$ in Abell 262. {Middle panel:} [OII] emission velocity dispersion map. The magenta line outlines the region where ALMA CO (2-1) emission is detected. {Right panel:} VSFs of nebular gas with different [OII] brightness levels in Abell 262. The red points show the VSF of nebular gas within the central region where ALMA CO emission is observed. The VSF results for the top 25\% brightest [OII] pixels are shown with black squares, while those for the top 50\% brightest [OII] regions are indicated by green squares.}
    \label{fig:a262_vsf_compare}
\end{figure*}

To further explore the motion of nebular gas in the core of Abell 262, we compare the VSFs of different [OII] brightness components within the central region where ALMA CO emission is observed. The continuum-subtracted [OII] flux map, presented in the left panel of Figure~\ref{fig:a262_vsf_compare}, reveals that the pixels with the highest 25\% brightness in [OII] emission are concentrated near the nucleus of Abell 262, surrounded by the outer top 50\% brightness component. The [OII] emission velocity dispersion map, illustrated in the middle panel of Figure~\ref{fig:a262_vsf_compare}, reveals a central high dispersion region extends along the axis of the AGN-driven radio jets. This suggests that the enhanced velocity dispersion may be a result of the jets colliding with the rotating gas disk \citep{gingras2024complex}.

The right panel of Figure~\ref{fig:a262_vsf_compare} shows that the VSFs of all [OII] brightness components exhibit identical slopes of $\sim$1, suggesting coherent rotation around the center. The region containing the top 25\% brightest [OII] emission has the highest VSF amplitudes, shown as the black points. A plausible explanation for the observed variation in VSF amplitudes is the differential rotation of ionized gas around the cluster's center, with higher angular velocities at the center contributing to the higher mean velocity differences observed in the brightest [OII] regions. Another possible reason is the projection effect, which will be discussed in detail in Section~\ref{sec:projection}. The central 25\% of the brightest [OII] emission is located in the high-density environment of the cluster core. More gas velocity components are aligned along the line of sight, which is consistent with the high velocity dispersion and contributes to the elevated VSF amplitudes measured.

\subsubsection{RXJ0820.9+0752}

RXJ0820 contains one of the most gas-rich central galaxies known. However, as seen in the bottom left panel of Figure~\ref{fig: keck_v50_maps}, unlike the rest of the sample in this study, the [OII] line emission is not centered on the cluster nucleus but is instead offset to the northwest of the central galaxy, extending 24 kpc in RA and 15 kpc in DEC. {A secondary galaxy located at $\sim$8 kpc SE to the main galaxy, moves at a relative velocity of $\rm \sim 100~km~s^{-1}$ \citep{kim2002peculiar,olivares2019ubiquitous}. }Therefore, it is plausible that sloshing motions, induced by the interaction with this nearby galaxy, might also contribute to the formation of and significant offset of cold gas reservoirs in RXJ0820 \citep{vantyghem2019enormous}.

The median velocity of [OII] emission is substantially redshifted compared to the central stellar velocity, ranging between $-12~\rm km~s^{-1}$ and $+286~\rm km~s^{-1}$. The region with the highest median velocity is located approximately \(4 \, \text{kpc}\) north of the nucleus, while the gas with the lowest velocities is found \(\sim 14 \, \text{kpc}\) west of the nucleus. Given that the nebular gas is largely undisturbed, with velocity dispersion below $100~\rm km~s^{-1}$, only a single Gaussian component is required to fit the velocity distribution of the ionized gas. This smooth velocity map, which lacks complex structures, is in a quiescent phase. 

\begin{figure*}
\centering
	\includegraphics[width=\textwidth]{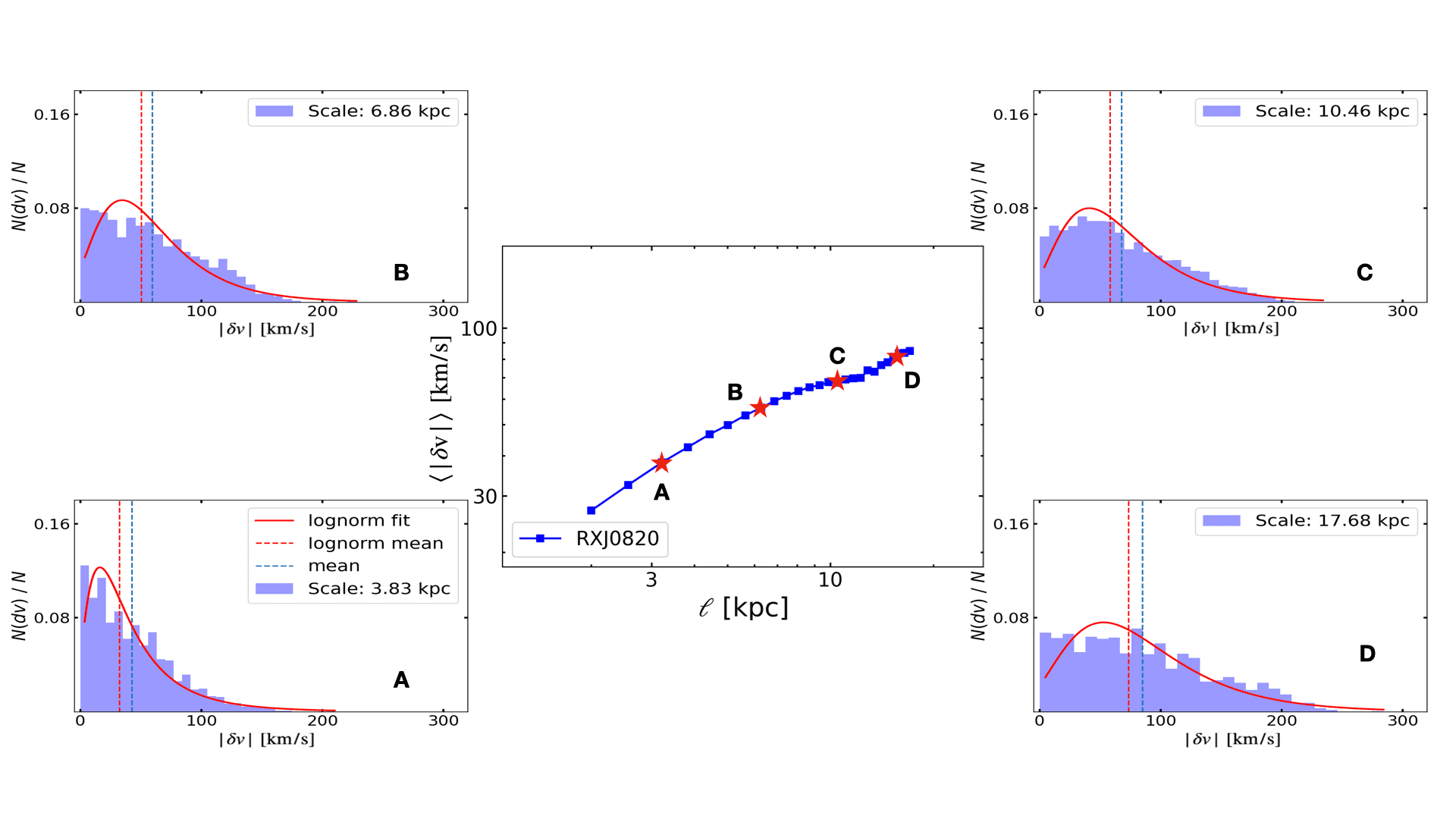}
 \caption{\textbf{Middle:} Similar to Figure~\ref{fig:keck_vfs_a1835}, the middle panel displays the VSF for the nebular gas within the core of RXJ0820.9+0752. {Side panels:} the distributions of $|\delta v|$ within four separation bins.}
    \label{fig:keck_vfs_rxj}
\end{figure*}

Figure \ref{fig:keck_vfs_rxj} presents the VSF for the nebular gas in the core of RXJ0820. The amplitude of the VSF increases from below $30~\rm km~s^{-1}$ to $85~\rm km~s^{-1}$ over the scales spanning from 2 to 17 kpc. The VSF is linear with a slope of  \(\sim 0.61\) on scales of 2 - 8 kpc. The slope flattens progressively at larger scales around 9 kpc, which is close to the size of X-ray bubble in RXJ0820, suggesting that the central AGN activity may drive the motion of nebular gas. 
The log-normal distribution can adequately describe the \(|\delta v|\) distributions within the measured scales, as shown by the red solid lines in the side panels of Figure~\ref{fig:keck_vfs_rxj}. This suggests that the high \(|\delta v|\)  tails, which are likely due to projection effects on small scales, can be mitigated by using a log-normal fit.

\subsubsection{PKS 0745-191}
\label{sec:pks0745}
The bottom right panel of Figure \ref{fig: keck_v50_maps} displays the nebular median velocity map of the core of PKS 0745. The nebular gas shows an average redshift of $\rm \sim 135~km~s^{-1}$ relative to the systemic stellar velocity, ranging from -75 $\rm km~s^{-1}$ to +400 $\rm km~s^{-1}$. PKS 0745 has the broadest emission lines among four central galaxies, indicative of a highly disturbed environment with velocity dispersion up to $\rm 400 ~km~s^{-1}$. As a result, the line profiles of about 16\% of the spaxels need to be modeled with two Gaussian velocity components. The dynamics of each gas component will be further discussed in Section \ref{sec:components}. The central galaxy of PKS 0745 contains a powerful radio source, emitting energy ten times higher than the other targets \citep{pulido2018origin}. The X-ray cavities, with diameters $\sim$ 10 kpc and 17 kpc and located approximately 6 kpc and 20 kpc from the center along the NW - SE axis, are asymmetrically positioned relative to the nucleus \citep{sanders2014feedback}, closely following the [OII] emission distribution, which spans 33 kpc in RA and 30 kpc in DEC. The core of PKS 0745 contains $4.9 \times 10^9 M_\odot$ of cold molecular gas, observed by ALMA CO (3-2) emission line within 5 kpc of the nucleus \citep{russell2016alma}, as indicated by the central black contours. It is mostly concentrated in the brightest nebular emission area and lies in filaments trailing behind the X-ray cavities in the system.

\begin{figure*}
\centering
	\includegraphics[width=\textwidth]{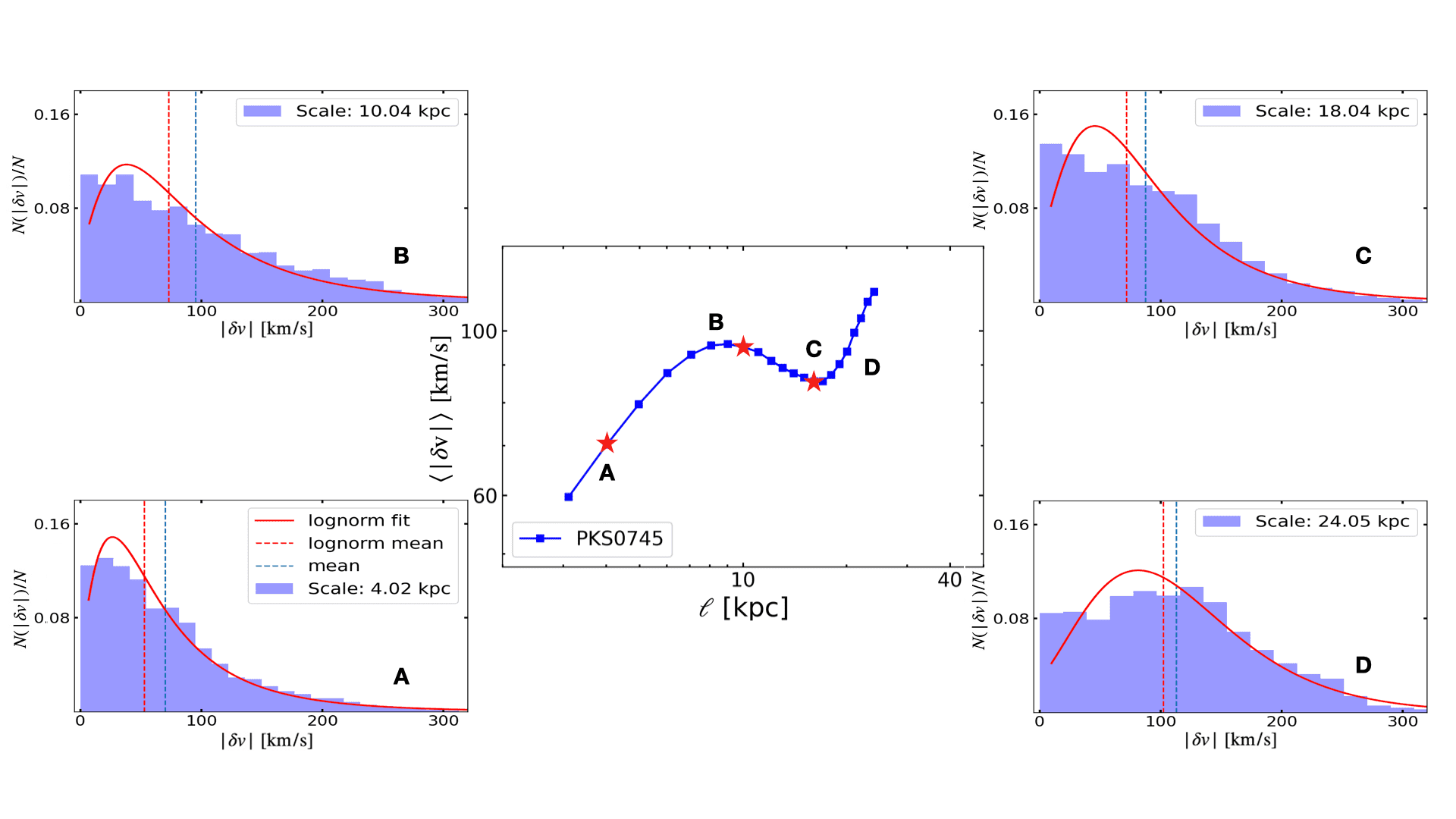}
    \caption{\textbf{Middle:} Similar to Figure~\ref{fig:keck_vfs_a1835}, the middle panel displays the VSF for the nebular gas within the core of PKS 0745-191. {Side panels:} the distributions of $|\delta v|$ within four separation bins.}
    \label{fig:keck_vfs_pks}
\end{figure*}

Figure \ref{fig:keck_vfs_pks} illustrates the VSF of the nebular gas in PKS 0745. Starting from the smallest measured scales of 3 kpc to 8 kpc, the VSF rises linearly from around $\rm 60 ~km~s^{-1}$ to a peak of approximately $\rm 95~ km~s^{-1}$ with a slope of 0.50. {The VSF displays a turnover at a scale of $\sim 9$ kpc, which is comparable to the size of the innermost cavity, which is $6$ kpc southeast to the nucleus.} It then drops to a minimum of $\rm 85~ km~s^{-1}$ at $\sim$15 kpc before rising again to over $\rm 120~ km~s^{-1}$ at the largest measured scale of 24 kpc. The limited FOV restricts the measurement of VSFs at larger scales. However, the complex VSF characteristics observed in PKS 0745 may indicate that the gas motions are driven by multiple cycles of central AGN activities. In side panels, all four $|\delta v|$ distributions exhibit left-skewed peaks with extended tails, and are well modeled by log-normal profiles, similar to those observed in RXJ0820 and Abell 1835.

\subsection{VSF of the two velocity components in Abell 1835 and PKS 0745-191}

\label{sec:components}

Emission line velocity profiles reveal the gas kinematics. A single Gaussian component often inadequately represents the line profile in the presence of multiple kinematic components. Two Gaussian velocity components were detected in Abell 1835 and PKS 0745 \citep{gingras2024complex}, described in Section~\ref{sec:keck_data_reduction}. The first Gaussian component, representing the “core” of the emission line, is typically narrower than the second component, which often has a broad wing at the base of the emission line. 
Multiple gas complexes along the line of sight can contribute to the line broadening. Turbulence introduces random motions resulting in a broadened velocity distribution. Additional broadening may arise from bulk motions associated with outflows, inflows, dynamical stirring, activities of supernovae and AGN. To understand the kinematics of each component in Abell 1835 and PKS 0745, we calculate the VSFs of both velocity components of the ionized gas individually.

\begin{figure*}
\centering
    	\includegraphics[width=17.5cm,height=11cm]{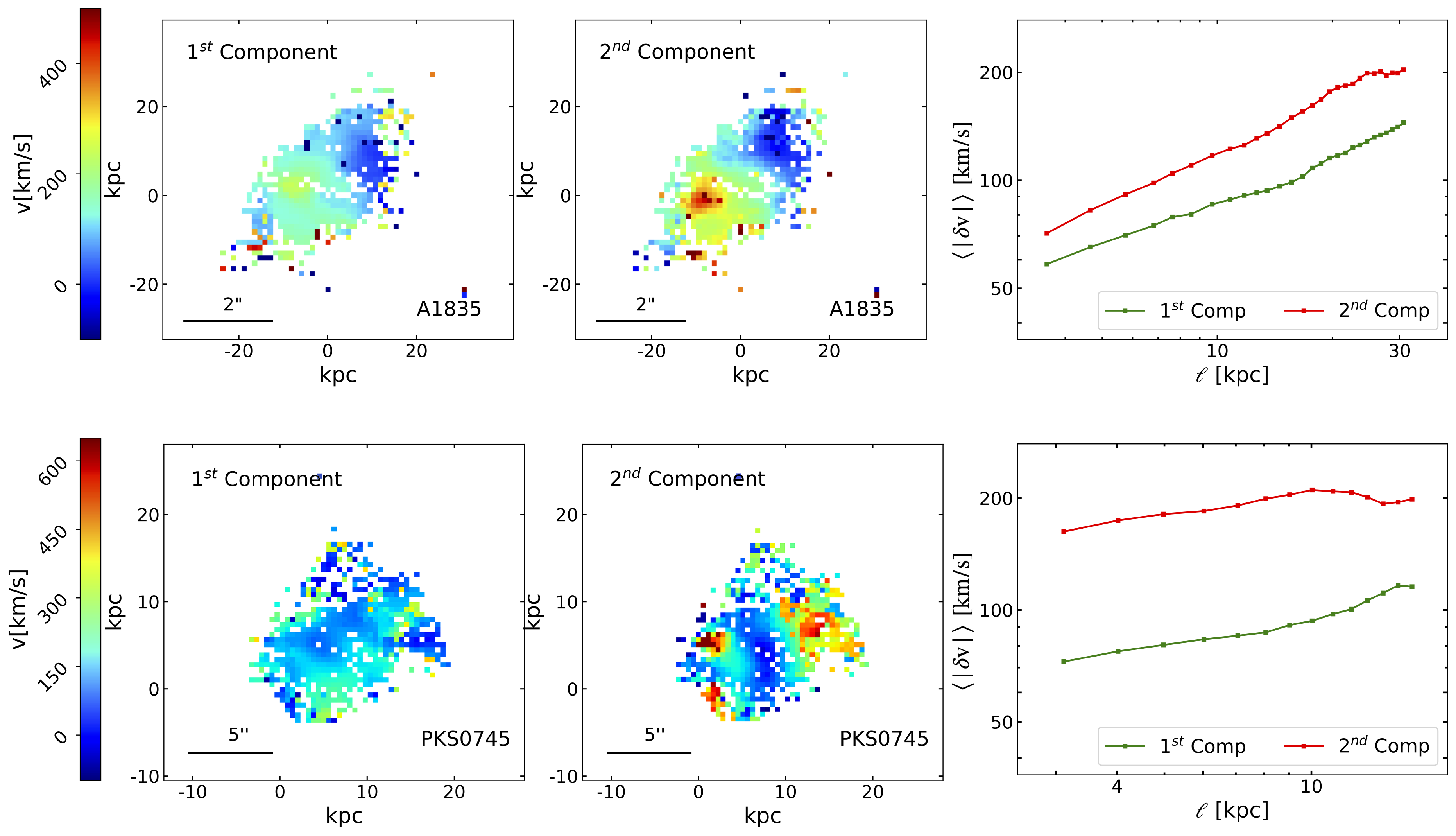}

    \caption{Velocity maps and VSFs of the first Gaussian component and the second Gaussian component derived from KCWI [OII] emissions in the core of Abell 1835 (top row) and PKS 0745 (bottom row). The maps include only spaxels where emission lines can be accurately fitted by two kinematic components with a (S/N) $\ge$ 3. In the right panels, the VSFs for the 1st and 2nd components are represented by green and red points, respectively.}
    \label{fig:a1835_pks_component}
\end{figure*}

The left and middle panels of Figure~\ref{fig:a1835_pks_component} show the median velocity of the first and second Gaussian kinematic components in the core of Abell 1835 (top row) and PKS 0745 (bottom row), respectively. Only pixels with both components detected at a significance of  \( \text{S / N } > 3 \) are presented. The narrow component likely reflects the more extensive outer, quiescent region of the nebula, while the broad component, which is embedded within the narrow component, is primarily associated with the central region near the radio bubbles in Abell 1835 and PKS 0745-191 \citep{gingras2024complex}.   

The second kinematic component exhibits a significantly larger velocity variance across the field than the quiescent first component. A velocity gradient is observed in the velocity map of the second component along the NW-SE axis of the X-ray cavities in Abell 1835 \citep{gingras2024complex}. Additionally, highly redshifted gas with velocities exceeding 500 $\rm km~s^{-1}$ is detected to the NW, E, and SE of the nucleus in the velocity map of the second component for PKS 0745, following the positions of X-ray cavities.

\begin{deluxetable*}{lccccc}
\tablecaption{The VSFs of two velocity components in Abell 1835 and PKS 0745-191\label{tab:component_vsf}}
\tablewidth{0pt}
\tablehead{
    \colhead{Target} & 
    \colhead{Component} & 
    \colhead{Scales} & 
    \colhead{Turnover Scale} & 
    \colhead{Scales for Fit} & 
    \colhead{Slope} \\
    \colhead{} & 
    \colhead{} & 
    \colhead{(kpc)} & 
    \colhead{(kpc)} & 
    \colhead{(kpc)} & 
    \colhead{}
}
\decimalcolnumbers
\startdata
Abell 1835 & 1$^{st}$ & 4 - 30 & -- & 4 - 17 / 17 - 30 & 0.35 / 0.53 \\
           & 2$^{nd}$ & 4 - 30 & 23 & 4 - 23 & 0.53 \\
PKS 0745-191 & 1$^{st}$ & 3 - 15 & -- & 3 - 8 / 8 - 15 & 0.19 / 0.43 \\
             & 2$^{nd}$ & 3 - 15 & 10 & 3 - 10 & 0.21 \\
\enddata
\tablecomments{This table shows the VSFs for the first and second Gaussian components detected by the [OII] emission in Abell 1835 and PKS 0745-191. The scales indicate the range of spatial extents studied, the turnover scale where the VSF changes behavior, and the range used for fitting the slopes of these functions.}
\end{deluxetable*}

The right panels of Figure~\ref{fig:a1835_pks_component} show the VSFs for the first and second Keck [OII] components in Abell 1835 and PKS 0745. The second component, indicated by the red points, consistently shows higher VSF amplitudes across all scales, while the VSFs of the first component, represented by the green points, exhibit lower amplitudes, reflecting its milder variance compared to the second component. A summary of the VSF measurements is provided in Table~\ref{tab:component_vsf}. 

In Abell 1835, both velocity components exhibit approximately linear growth as the scale increases, with a steeper progression observed at larger scales, approaching a slope of 1/2. This steepening at larger scales may be attributed to the gravitational acceleration of the gas, as discussed in more detail in Section~\ref{sec:steepening}. Additionally, the VSF of the second component begins to flatten beyond the $\sim$ 23 kpc scale, which corresponds to the distance of the X-ray bubbles from the center of Abell 1835. {In A1853, the VSFs of [OII] emission obtained using both components, as presented in Figure 3, also exhibit a turnover at this scale. It suggests that the scale of energy injection into the gas motions is closely linked to the rising bubbles and the dynamics of the turbulent component.}

A similar pattern for both components is observed in PKS 0745. The first component shows comparable linear growth with a slope of $\sim$0.2 at scales below 8 kpc, followed by a steeper increase of $\sim$0.43. Meanwhile, the VSF of the second component, represented by the red points, exhibits a significantly higher amplitude. A turnover is observed at the scale of 10 kpc, which is close to the size of the secondary southeast X-ray cavity in PKS 0745. {A turnover is also observed at the same scale in the VSFs of PKS0745, which uses a double-component V50, as shown in Figure \ref{fig:keck_vfs_pks}. This observation further supports the idea that gas motions are induced by AGN activity.}

Overall, these findings suggest that the two different Gaussian components trace two distinct kinematic components with different origins. The second component likely reflects the turbulent motion of gas driven by AGN activity in the galaxy. Additionally, gravity likely also has a significant influence on the motion of both components.


\subsection{Molecular Gas Distribution and Kinematics}

\label{sec: alma vsf}

Here we compare the kinematics of the nebular gas to the cold molecular gas. Figure~\ref{fig: alma_v_maps} shows the line of sight velocity maps of the dominating component of CO emission observed by ALMA in our samples. The CO emission distribution is filamentary, clumpy, and asymmetric. Most of the gas lies in extended filaments outside of the nucleus surrounding or extending towards X-ray cavities inflated by radio jets.

In PKS 0745, {$\rm 4.6 \pm 0.3 \times 10^9~M_\odot$ }molecular gas is observed in three filaments extending in the N, SW, and SE directions \citep{russell2016alma}. The N and SW molecular filaments extend up to 5.7 kpc and 3.6 kpc, respectively, towards the X-ray cavities. Abell 1835 contains $\rm 5\times10^{10}~M_{\odot}$ of molecular gas, detected by CO(3-2) and CO(1-0) emission with ALMA \citep{mcnamara2006starburst}. The molecular gas is concentrated around the central galaxy and three filaments extending \(35 \, \text{kpc}\) in RA and $\sim$\(25 \, \text{kpc}\) in DEC. Two filaments, projecting towards the NW and SE X-ray cavities, appear to be part of a bipolar outflow with a mass of around $\rm 10^{10}~M_{\odot}$, which may be driven outwards by the mechanical energy from the buoyantly rising bubbles. 

In contrast, the molecular clouds in RXJ0820 are displaced northward 3 kpc from the nucleus, similar to the nebular gas. The cold gas is concentrated into two bright clumps. The brighter clump is approximately 3 kpc north of the BCG center, while the secondary clump is 4.6 kpc west of the primary clump. The cavity is too feeble to lift enough low-entropy gas to explain the observed gas via simulated cooling. Instead, the sloshing motions in the ICM, induced by the close passage of a nearby galaxy, might contribute to the condensation of the cold gas and its displacement \citep{vantyghem2019enormous}. Abell 262 is the only system here with a gaseous, circumnuclear disk approximately 3 kpc across, whose angular momentum axis lies approximately perpendicular to the radio jets and X-ray bubbles \citep{gingras2024complex}.

\begin{figure*}
\centering
	\includegraphics[width=16cm,height=14cm]{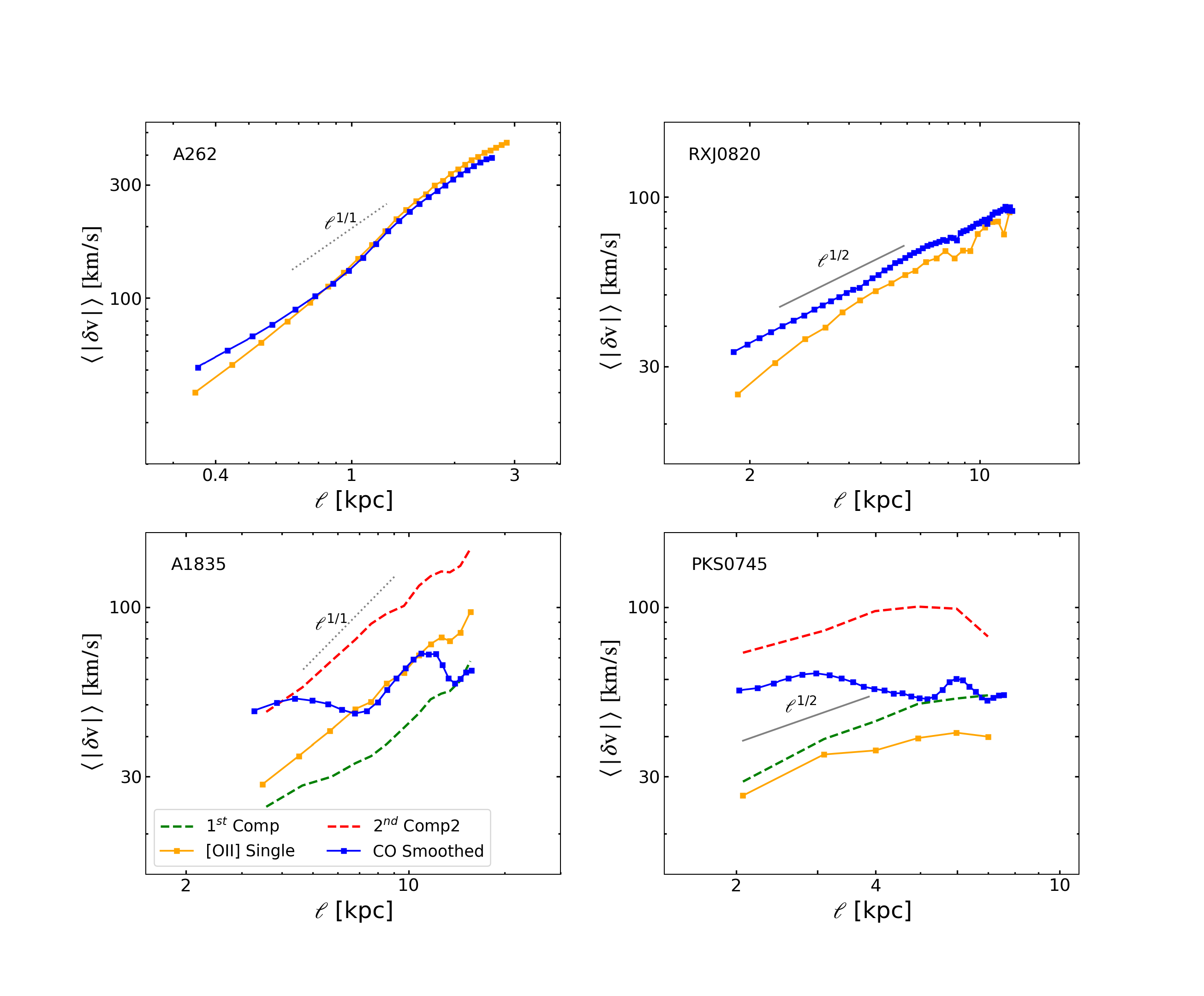}
    \caption{Comparison of the VSFs of cold molecular gas and warm ionized gas in our sample clusters, traced by ALMA CO emission and Keck [OII] line emission, respectively. The VSFs of the single component v50 (yellow points), the 1st (green dashed line) and the 2nd (red dashed line) velocity component from the Keck observations are computed within the regions covered by ALMA CO emission. 
    For each cluster, the ALMA velocity map is smoothed with a box kernel to match the ALMA observation beam size with the KCWI seeing, and the smoothed CO emission VSF is shown as blue points. Only scales exceeding the ALMA observation beam size and KCWI seeing limits, and having a sufficient number of pixel pairs for reliable VSF calculation, are displayed here. Reference lines with slopes of 1/1, and 1/2 are also included.}
    \label{fig:alma_keck_vsf_comparison}
    \end{figure*}

To investigate the relationship between different gas phases and their roles in AGN feedback, we calculate the VSF of the cold gas observed by ALMA and compare it with that of the warm ionized gas, as illustrated in Figure~\ref{fig:alma_keck_vsf_comparison}. Here the median velocity of the Keck [OII] emission is modeled using a single Gaussian component for all four targets, including Abell 1835 and PKS 0745, as the velocity of the molecular gas is similarly characterized by a single dominant component. To ensure accuracy, we focused only on central regions where both CO and [OII] emissions are detected, as indicated by the black dashed lines in Figure~\ref{fig: keck_v50_maps}. The 
 yellow points represent the VSF of warm nebular gas. Due to the compact nature of the CO emission, the measurements are confined to scales of approximately 10 kpc in all targets. A measurement summary is provided in Table~\ref{tab:co_oii_vsf_comparison}.

We examined and accounted for the systematic effects of telescope resolution, seeing, and smoothing applied during data processing to the comparison of VSFs obtained from multiple instruments. Smoothing determines the low separation cut-off, the VSF slope, and amplitude. It decreases the velocity differences across the field, leading to amplitude suppression, particularly at scales below the smoothing kernel width, further steepening the slope. The smoothing effect on VSF is detailed in Appendix~\ref{appdenix:smoothing}.

The KCWI data has a lower spatial resolution than ALMA. Its smallest spatial scale probed for VSF measurement is determined by the seeing for each object. The ALMA observations are smoothed by their synthesized beam sizes during data reduction. 
For a direct and meaningful comparison, we smoothed the ALMA maps using box kernels to match their synthesized beam size to the seeing of the KCWI observations. Due to the asymmetry of the synthesized beam, we applied smoothing based on the length of its major axis. The VSFs of the smoothed CO emission are presented as the blue points in Figure~\ref{fig:alma_keck_vsf_comparison}. The unsmoothed CO emission VSFs are not shown, but the measurements are summarized in Table ~\ref{tab:co_oii_vsf_comparison}.

The VSF of the ALMA CO (2-1) emission in Abell 262 ranges from 39 $\rm km~s^{-1}$ at a scale of 0.2 kpc to 397 $\rm km~s^{-1}$ at a scale of 2.6 kpc, with a slope of $\sim$ 0.97. The smoothed ALMA VSF slope increases to $\sim$ 1.12, as shown by the blue points in the upper right panel of Figure~\ref{fig:alma_keck_vsf_comparison}, nearly aligning with the KCWI VSF of 1.21. Despite slight differences observed at smaller scales, the molecular gas VSF exhibits similar amplitudes to that of the nebular gas, represented by the yellow points. This consistency suggests that the warm and cold gas phases are dynamically coupled in their rotation around the nucleus.

The cold molecular and nebular gases in RXJ0820 also exhibit a close coupling. The smoothed VSF of CO (1–0) emission ranges from 33 $\rm kms^{-1}$ to 91 $\rm kms^{-1}$ at scales of 1.8 to 12.5 kpc, with a slope of $\sim$ 0.53. The KCWI [OII] VSF within the CO emission region exhibits a slightly steeper slope of 0.62 but with lower amplitudes, particularly at smaller scales. Nonetheless, both phases increase similarly as scales increase and show no turnover. Considering both cold molecular and nebular gas offset from the cluster center, although mild differentiation between the two gas phases is seen, their kinematics, overall, are similar.

Compared to the quiescent gas in Abell 262 and RXJ0820, the kinematics of the churned-up gas in Abell 1835 and PKS 0745 are more complex. The KCWI [OII] emission VSF of Abell 1835 increases smoothly with a slope of 0.85, while the smoothed molecular gas VSF shows a turnover at scales of approximately 13 kpc, which is close to the size of the X-ray cavity in Abell 1835. Due to the limited extent of the detected molecular gas, the VSF cannot be measured at scales beyond 16 kpc. Consequently, the turnover at larger scales may be influenced by the window effect, as discussed in Appendix~\ref{appdenix:window}. The molecular and nebular gases are tightly coupled on scales between 7 and 13 kpc. However, the gas phases begin to differentiate more at smaller and larger scales. The correlated kinematics at intermediate scales may reflect similar morphologies along the bubble axis where the outflow is observed \citep{gingras2024complex}.

The velocity structure of the CO (3-2) line emission, resolved into three main filaments that contain roughly 90 percent of the total molecular gas mass in the BCG of PKS 0745, is used to compute the VSF of the cold phase gas, and compared to the [OII] emission in the bottom right panel of Figure~\ref{fig:alma_keck_vsf_comparison}. The VSF of the smoothed ALMA CO emission, shown as the blue points, varies from 52 $\rm km~s^{-1}$ to 62 $\rm km~s^{-1}$ at scales of 2.0 to 7.6 kpc, which is lower than that measured in the other three targets. This is consistent with the low-velocity dispersion of less than 150 $\rm km~s^{-1}$, which is significantly lower than the typical stellar velocity dispersion of such a massive BCG. The [OII] emission VSF differs in amplitude from the molecular gas, and at some points has opposite signs, indicating the two gas phases are moving separately. Due to the limited extent of CO emission, we are unable to measure its VSF on the characteristic scales of bubbles in this system.

The comparison between the first and second velocity components of [OII] emission in Abell 1835 and PKS 0745 and the dominant component of molecular gas, as shown in Figure~\ref{fig:alma_keck_vsf_comparison}, presents both intriguing insights and challenges for interpretation. The first (green dashed line) and second [OII] (red dashed line) components in both systems exhibit slopes similar to those of the single Gaussian VSFs but differ in amplitude by more than a factor of two. Notably, these components appear to be disconnected from the molecular gas, a phenomenon that remains difficult to understand. A major factor may be sensitivity, as nebular emission is more sensitive to gas mass by two to three orders of magnitude. Therefore, more sensitive ALMA observations would be required to probe these components at the level of detail provided by the [OII] observations.

In summary, the nebular and molecular gas VSFs for Abell 262 and RXJ0820, apart from mild differentiation, are broadly consistent indicating strongly coupled kinematics on all scales. However, the VSFs for PKS 0745 and Abell 1835, which are experiencing strong radio-AGN feedback, are more complex. The gas is correlated only at some scales in Abell 1835 but is differentiated at all scales in PKS 0745.

\begin{deluxetable*}{lcccccccccccc}
\tablecaption{Comparison between the VSFs of cold molecular gas and warm ionized gas \label{tab:co_oii_vsf_comparison}}
\tablewidth{\textwidth}
\tabletypesize{\scriptsize}
\tablehead{
\colhead{Target} & \multicolumn{4}{c}{CO emission} & \multicolumn{4}{c}{Smoothed CO emission} & \multicolumn{4}{c}{[OII] emission} \\
\nocolhead{} & \colhead{Scales} & \colhead{$|\delta v|$} & \colhead{Scale for fit} & \colhead{Slope} &
\colhead{Scales} & \colhead{$|\delta v|$} & \colhead{Scale for fit} & \colhead{Slope} &
\colhead{Scales} & \colhead{$|\delta v|$} & \colhead{Scale for fit} & \colhead{Slope} \\
\nocolhead{} & \colhead{(kpc)} & \colhead{($\rm km~s^{-1}$)} & \colhead{(kpc)} & \nocolhead{} &
\colhead{(kpc)} & \colhead{($\rm km~s^{-1}$)} & \colhead{(kpc)} & \nocolhead{} &
\colhead{(kpc)} & \colhead{($\rm km~s^{-1}$)} & \colhead{(kpc)} & \nocolhead{} \\
\colhead{(1)} & \colhead{(2)} & \colhead{(3)} & \colhead{(4)} & \colhead{(5)} &
\colhead{(6)} & \colhead{(7)} & \colhead{(8)} & \colhead{(9)} &
\colhead{(10)} & \colhead{(11)} & \colhead{(12)} & \colhead{(13)}
}
\startdata
Abell 1835 & 1.4 - 15.8 & 36 - 77 & 6.8 - 11.6 & 0.83 & 3.3 - 15.8 & 48 - 72 & 6.8 - 11.6 & 0.92 & 3.3 - 15.6 & 28 - 97 & 3.3 - 15.6 & 0.85 \\
Abell 262  & 0.2 - 2.6  & 39 - 397 & 0.2 - 2.6 & 0.97 & 0.4 - 2.6 & 55 - 391 & 0.4 - 2.6 & 1.12 & 0.4 - 2.8 & 40 - 452 & 0.4 - 2.8 & 1.21 \\
RXJ0820.9+0752 & 0.5 - 12.5 & 21 - 98 & 0.5 - 12.5 & 0.48 & 1.8 - 12.5 & 33 - 91 & 1.8 - 12.5 & 0.53 & 1.8 - 12.3 & 25 - 90 & 1.8 - 12.3 & 0.62 \\
PKS 0745-191 & 0.6 - 7.6 & 48 - 71 & / & / & 2.0 - 7.6 & 52 - 63 & / & / & 2.4 - 7.3 & 30 - 43 & 4.2 - 6.8 & 0.43 \\
\enddata

\tablecomments{Columns: (1). Target. (2) - (5). VSFs of CO emission: scales can be measured, measured VSF, scales for slope fitting, best-fit slope. (6)-(9). VSFs of smoothed CO emission. (10) - (13). VSFs of [OII] emission in CO emission detected region.}
\end{deluxetable*}


\subsection{Velocity Spectra of Hot ICM}
\label{sec:3.5}
We use existing Chandra X-ray observations to estimate the velocity spectra of hot ICM motion in our targets. The X-ray emissivity in the soft energy band (0.5 - 3.5 keV is used in this work) is weakly dependent on temperature when the electron temperature exceeds 3 keV. Therefore, the SB fluctuations observed in X-ray images reflect gas density fluctuations. First, the ‘unperturbed’ SB gradient is removed by dividing the raw images by the best-fit spherically symmetric $\beta$-model. Then, the rms of SB fluctuations at wavenumber \(k\) in the residual images are calculated using the modified $\Delta$-variance method \citep{arevalo2012mexican}, and are deprojected to derive the scale-dependent amplitudes of gas density fluctuations, $\rm A_{3D}(k)$ \citep{churazov2012x}.

Using cosmological simulations of relaxed galaxy clusters, \cite{zhuravleva2014relation} provide a theoretical argument and confirm the linear scaling relation between the amplitudes of gas density fluctuations and velocity fluctuations across a wide range of scales:

\begin{equation}
A_{3D}(k) \equiv \frac{\partial \rho_k}{\rho_{0}} = \eta_1 \frac{V_{1,k}}{c_s},
\end{equation}
where \(c_s\) is the sound speed in the atmosphere, and \(\eta_1 \approx 1 \pm 0.3\) is the proportionality coefficient. The velocity spectra $V_{1,k}$ in this work are defined as velocity fluctuation at scale $l\equiv k^{-1}$, rather than the energy distribution of velocity fluctuations obtained from the Fourier transform of the velocity field. This method overcomes the challenge of directly measuring gas velocity due to the low spectral resolution of current X-ray observations. The indirect velocity spectra measurements \citep{zhuravleva2018gas} are consistent with the direct measurements of line-of-sight velocity dispersion obtained by Hitomi \citep{hitomi2016quiescent} within the central 30  - 60 kpc of the Perseus cluster.

\begin{figure*}
     \centering
     \includegraphics[width=17.5cm,height=5.5cm]{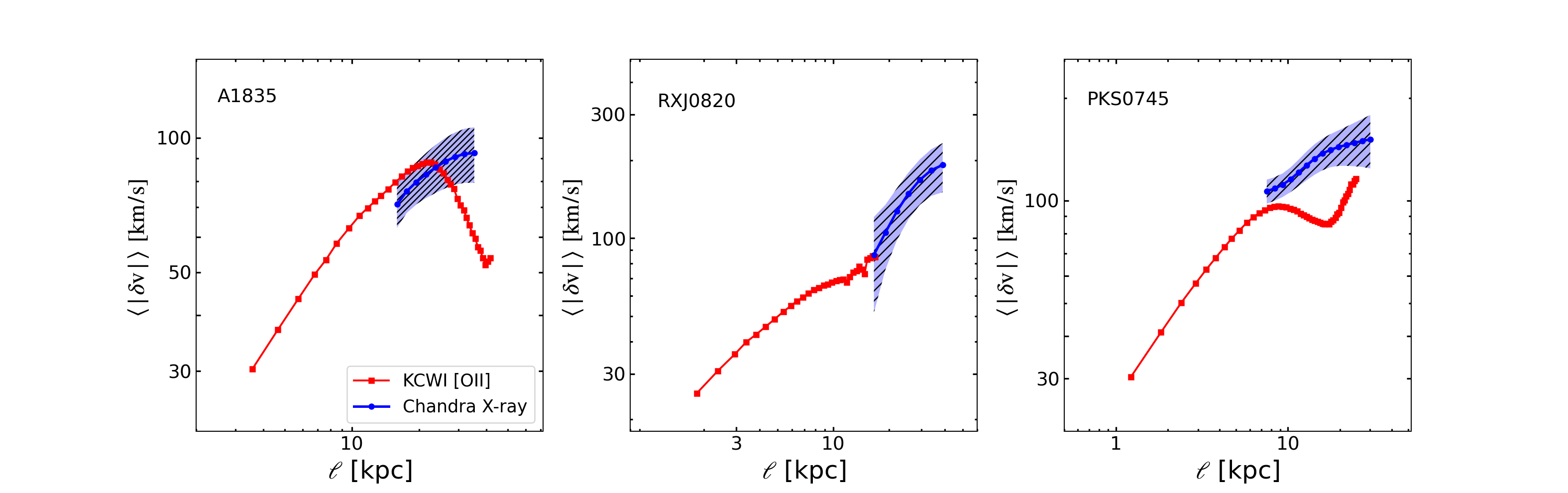}
    \caption{Comparison between the VSFs of warm ionized gas traced by Keck [OII] line emission and the Chandra X-ray surface brightness fluctuation analysis in the cool cores of Abell 1835, RXJ0820.9+0752, PKS 0745-191. The VSFs of warm ionized gas are presented as the red points. Only scales exceeding the seeing limits, and with a sufficient number of pixel pairs for reliable VSF calculation, are displayed here. The power spectra of hot ICM motion are computed within the regions covered by KCWI [OII] emission and are depicted by the blue points. The hatched regions reflect $1\sigma$ uncertainty. Only scales least affected by systematic uncertainties and Poisson noise are shown. The X-ray results are summarized in Table~\ref{tab:sb_analysis_vk}}
    \label{fig:chandra_keck_vsf}
    \end{figure*}

 The projected areas where CO emission is detected are too small for X-ray SB analysis, therefore, a direct comparison to cold molecular gas is not included in this work. The velocity amplitudes of hot atmosphere motions are estimated within the KCWI [OII] emission detected regions, marked as blue points in Figure~\ref{fig:chandra_keck_vsf}. The width of the blue regions reflects the $1\sigma$ statistical uncertainty. Due to the predominance of Poisson noise at small scales, we only present measurements at scales where the power of X-ray SB fluctuations is larger than that of the Poisson noise. Deeper observations are needed to accurately constrain gas perturbations on smaller scales, allowing us to compare multi-phase gas dynamics across a wider range of scales. Table~\ref{tab:sb_analysis_vk} lists the X-ray measurements in each cluster. Given the similar rotational kinematics observed between the molecular and nebular gas in the core of Abell 262, both exhibit clear signs of ordered motion, which is beyond the limitations of X-ray SB analysis. 
  Therefore, we focused the analysis on the other three targets, excluding Abell 262.

    The right panel of Figure~\ref{fig:chandra_keck_vsf} shows the one-component velocities of hot gas motions observed within the [OII] emission-detected area in the core of {Abell 1835, ranging from 71 $\rm km~s^{-1}$ at 18 kpc scale to 92 $\rm km~s^{-1}$ at a 35 kpc scale.} 
    Although the energy injection scale cannot be constrained from the SB fluctuation analysis, the velocity spectrum and VSF of [OII] emission show consistency around the [OII] VSF turnover scales of $\sim 20$ kpc, close to the bubble size in this system. Combined with the cold gas distribution along the cavity axis, it suggests that the nebular filaments likely condensed from the turbulent hot ICM due to thermal instability, triggered by the inflation of X-ray bubbles related to central AGN activities in Abell 1835.

The velocity spectrum of RXJ0820 
ranges between 68 $\rm km~s^{-1}$ at a 17 kpc scale and 192 $\rm km~s^{-1}$ at a 40 kpc scale, shown in the middle panel of Figure~\ref{fig:chandra_keck_vsf}. The FOV of [OII] emission observations limits direct comparison between the VSF of the nebular gas and the X-ray velocity spectrum at large scales. Nevertheless, at the 17 kpc scale, where overlap occurs, the VSF amplitude of the nebular gas and the derived velocity of hot atmospheric motion are both measured at approximately 80 $\rm km~s^{-1}$, indicating their potential coupling. Given that the motion of cold filaments is also closely coupled with that of the nebular gas, there are two possible scenarios in which the filaments could share similar turbulent motions with the hot ICM: (1) the filaments form from the hot atmospheres, allowing them to retain the turbulent characteristics of the hot gas, or (2) the filaments, even though formed independently, evolve to be co-moving with the hot gas over time. 

In the right panel of Figure~\ref{fig:chandra_keck_vsf}, {the hot atmosphere in PKS 0745 exhibits a velocity spectrum ranging between 107 $\rm km~s^{-1}$ and 151 $\rm km~s^{-1}$ at scales of 8 kpc to 30 kpc.} The VSF of the nebular emission is lower than that of the hot ICM, and the turnover at the scale of 9 kpc seen in the ionized gas is not observed in the X-ray measurements.

The slope of the velocity spectra may be affected by the 
$\Delta$-variance method used in the X-ray surface brightness analysis. Furthermore, the considerable uncertainties presented in Figure~\ref{fig:chandra_keck_vsf} correspond to a broad spectrum of slopes, preventing us from confirming or ruling out Kolmogorov turbulence. Simulations including magnetic stresses in \cite{fournier2025vsf} found no significant correlation between the behavior of the hot and cold phases of model cool-core atmospheres. The VSF power-law index of the cold phase varies while the hot phase is more stable. The VSF amplitudes and indices vary significantly depending on viewing angle and atmospheric conditions, suggesting that these effects must be carefully considered when interpreting observed VSFs. 

In summary, in Abell 1835, RXJ0820, and PKS 0745, the velocity amplitudes of gas motion in the hot atmospheres determined by X-ray SB analysis range from $\rm 100~km~s^{-1}$ to $\rm 200~km~s^{-1}$, aligning with the expected turbulent velocities of the ICM. The cold and hot gas in Abell 1835 and RXJ0820, exhibit consistent kinematics at large scales, indicating a potential relationship in their formation and evolution. It suggests that the interaction and co-evolution of different gas phases play a significant role in the dynamics of galaxy cluster cores.

\begin{deluxetable}{lCCC}
\centering
\tablecaption{Velocity Spectra of Hot ICM\label{tab:sb_analysis_vk}}
\tablewidth{0pt}
\tabletypesize{\scriptsize} 
\tablehead{
\colhead{Target} & \colhead{$T$} & \colhead{Scales} & \colhead{$V_{\mathrm{turb}}$} \\
\nocolhead{} & \colhead{(keV)} & \colhead{(kpc)} & \colhead{(km s$^{-1}$)} \\
\colhead{(1)} & \colhead{(2)} & \colhead{(3)} & \colhead{(4)} 
}
\startdata
Abell 1835 & 2.7 & {16 - 35} & 71 - 92 \\
RXJ0820.9+0752 & 1.3 & 17 - 40 & 68 - 192 \\
PKS 0745-191 & 2.6 & {8 - 30} & 107 - 151 \\
\enddata
\tablecomments{Velocity spectra of gas motions derived from Chandra X-ray SB fluctuations analysis: (1). Cluster name, (2). Mean temperature in the region of interest, (3). Spatial scales least affected by Poisson noise,  (4). {Mean velocity amplitudes} of gas motions.}
\end{deluxetable}

\section{Discussion}

\label{sec:discussion}


\subsection{The VSF slope and steepening at small scales}

\label{sec:steepening}

 The VSFs of gas in all phases in Abell 262 exhibit clear characteristics of ordered motion, with a slope of $\sim$ 1, consistent with disk rotation. Therefore, our discussion focuses on the VSFs in the other three filament-dominated objects. In these clusters, the VSFs are steeper than the 1/3 slope of classic Kolmogorov turbulence in an incompressible fluid. We do not necessarily expect the VSF slopes of the warm and cold gas phases to follow the ideal Kolmogorov slope, which serves as a reference for comparison.

The VSFs of filamentary nebular structures with slopes $\geq 1/2$ are reported in the centers of Perseus, A2579, and Virgo \citep{2020ApJ...889L...1L}. {The VSFs of multiphase filaments in the centers of 10 galaxy clusters have been measured, and in all systems, the slopes are found to be steeper than those expected by classical Kolmogorov theory \citep{ganguly2023nature}.} Supersonic turbulence, which has a characteristic slope of $1/2$ has been seen in simulations of of galaxy clusters and AGN feedback, indicating that cold gas may fragment out of a supersonic outflow driven by AGN jets or winds, forming the extended filamentary nebulae \citep{qiu2020formation}. These supersonic turbulent structures can become ‘frozen’ in the cold gas and flatten rapidly over a short timescale of approximately 10 Myr \citep{2022ApJ...929L..30H}. Our observations reveal nebular and molecular gas speeds significantly below the hot atmospheres' sound speeds.

For warm ionized gas at a temperature of $\sim$ $10^4$ K, the sound speed is $\sim$ 10 $\rm km~s^{-1}$; thus, the VSF amplitudes measured are highly supersonic. However, the velocities we have observed pertain to external gas motions, and the filaments themselves do not inherently exhibit supersonic characteristics. Thus, it is still uncertain whether the steep VSF can be directly linked to the supersonic nature of the originating outflows.
 
Despite excluding measurements significantly affected by the smoothing effect, amplitude suppression persists across all ranges, not just contributing to the VSF steepening at smaller scales (as detailed in Appendix~\ref{appdenix:smoothing}). Other factors may also play a role.

Gravity will steepen the VSF. We investigated the influence of gravity following the methodology of \cite{wang2021non}: test particles with identical initial velocities fall from the same height, overshoot the center while decelerating outward and oscillate. We sample the particles over different time intervals and 
find that the VSF slopes are sensitive to the time interval sampled, with the slope approaching a maximum of $\sim$ 1 during the phase before the particles return toward the center. Over a longer trajectory, {which} includes several oscillation periods, the VSF of gravity-dominated motion flattens over time. However, it does not consistently converge to 1/2 or 1/3, as suggested by \cite{wang2021non}, but rather approaches zero.

If the cold filaments condense out of the hot atmosphere, they may retain the memory of the initial conditions of the hot gas and fall toward the center under gravity. However, they may not overshoot the center via ballistic motion and continue oscillating around it. The infalling filaments may collide with pre-existing cold gas clouds, leading to a reduction in their velocity or their disruption before reaching the center. The freefall of cold gas can also be slowed by {magnetic stresses, or velocity kicks due to bulk motion}. Therefore, during the infall process, the VSF slope of the cold-phase gas is likely to remain close to 1 or below. The steep slope due to gravity is more pronounced on larger scales because the velocity differences accumulate over greater displacements. This phenomenon may explain the decreasing differences in the VSFs of cold molecular gas and warm ionized gas at large scales, as observed in Figure~\ref{fig:alma_keck_vsf_comparison}.

Magnetic fields can stabilize gas motion by suppressing the growth of hydrodynamic instabilities through a mechanism known as ‘magnetic draping’. This stabilization can steepen the VSF of turbulence generated by AGN-driven bubbles, as the energy cascade to smaller scales is hindered. Additionally, magnetic fields introduce anisotropy into the velocity field, with more significant suppression along the field lines \citep{bambic2018suppression}.  \cite{mohapatra2022velocity} used simulations of homogeneous isotropic subsonic turbulence to study the VSF of multiphase gas. They found that in the hot phase, the gas remains more isotropic and is less constrained by magnetic tension forces. However, in the cold phase, magnetic pressure dominates, leading to a much steeper VSF, especially at smaller scales.

Some quasars exhibit Kolmogorov or shallower slopes. \citet{chen2023empirical} found that among the four quasi-stellar object (QSO) nebulae studied, only TXS0206-048's VSF follows the Kolmogorov slope, while the others are shallower. The flatter VSFs may indicate multiple energy injection scales \citep{zuhone2016mapping} including AGN, mergers, magnetic field stresses \citep{wang2021non}, or geometrical projection (see below in Section~\ref{sec:projection}). The conclusive interpretation is unclear, but the agents governing the VSF of the quasars in \citep{chen2023empirical} differ from the central galaxies studied here.

\subsection{Projection Effects}

\label{sec:projection}

After accounting for smoothing, the VSFs of nebular emission and molecular gas are broadly similar in the cores of Abell 262 and RXJ0820, indicating the coupling of gas kinematics. {However, the VSFs of [OII] emission, shown as yellow points in Figure~\ref{fig:alma_keck_vsf_comparison}, are consistently smaller and steeper than those derived from ALMA CO emission, shown as blue points, across all four targets, particularly at smaller scales.} Projection is likely responsible for these scale-dependent differences. Although the VSFs of gas in both phases are computed within the same projected region near the cluster center, the measured VSFs are influenced by the depth of the cloud along the line of sight $\rm L$. At a given scale $l<$ L, the VSF is expected to be steepened, and the intrinsic slope generally recovers at larger scales $l > $ L \citep{1987ApJ...317..686O,xu2020projected}. This effect is similar to the smoothing effect discussed in Appendix\ref{appdenix:smoothing} and is usually termed ‘projection smoothing’. The nebular gas envelops the cold molecular gas, thus having a larger depth and being more significantly affected by projection smoothing. It is consistent with the relatively large amplitude differences at small scales and the observed steepening of the KCWI VSF below the scale of $\sim$6 kpc in RXJ0820.

{The clearest example of a projection bias is seen in Abell 262, as shown in Figure~\ref{fig:a262_vsf_compare}. There the brightest [OII] emission has the highest velocity amplitude and is thus closer to the nucleus where the gas density is highest. The lower speeds of the fainter emission along any line of sight indicate emission further from the center where the gas density is lower. 

{The kinematics in Abell 1835 and PKS0745 are distinct from Abell 262. The large velocity differences observed at small scales cannot be fully explained by projection effects alone, indicating that the distinctions between the nebular and molecular gas are real.} Feedback acting on gases with significant variations in temperature, density, and volume filling factor may play a contributing role.

\citet{zhang2022bubble} found that projection in compact emission sources flattens the VSF rather than steepens it. The velocity difference distribution plots reveal the presence of high $|\delta v|$ tails, even at small scales. These pixel pairs may correspond to two physically distant points that appear close in projection, potentially leading to an overestimation of the VSF amplitude at small scales and thereby flattening the slope. 

{Both types of projection effects may occur simultaneously, but the resulting net effect depends on the distribution of clouds and the kinematics of their gas. We cannot recover the intrinsic slope by correcting for projection effects because we lack specific knowledge about these properties. In this study, we merely propose potential explanations for the discrepancies observed in the VSFs of [OII] and CO emissions across different scales, presented in Figure \ref{fig:alma_keck_vsf_comparison}.}


\section{Conclusions}
\label{sec:conclusion}

 The VSFs of [OII] emission are compared to cold molecular gas probed by CO emission along common lines of sight in four systems. In addition, the velocity spectra of the surrounding hot atmospheres are estimated from surface brightness fluctuations observed with the Chandra X-ray observatory. These comparative analyses enable us to explore correlations between gas phases from temperatures ranging between $30~\rm K$ and $10^7~\rm K$.  Several technical issues that can bias the interpretation of the VSF analyses are explored. The main conclusions are as follows:

\begin{itemize}
   \item The VSF of KCWI [OII] emission shows that the nebular gas is disturbed in three of four clusters. Abell 262 is the exception, where the gas is rotating about the nucleus.

  \item The cold molecular gas, traced by ALMA CO emission, is spatially aligned with [OII] emission in each central galaxy. A comparison of the VSFs in RXJ0820 and Abell 262, where the gas is either in rotation or relatively quiescent, shows similar motions across all scales, but with minor differences. {These similarities in motion may indicate common dynamics and perhaps a similar origin of formation.}

  \item In Abell 1835, the VSFs of cold and warm gas differentiate at small scales, while in PKS 0745, they differentiate across all scales. The differentiation may occur during interactions between the cold and warm media and the radio jets and X-ray bubbles. 

    \item In Abell 262, the VSF slopes of the [OII] emission and CO emission of $\sim$ 1, consistent with ordered motion in a ring or disk. In contrast, the VSFs in the remaining three systems are steeper than the classical Kolmogorov turbulence slope of 1/3. The steepening may be {caused} by a variety of factors including magnetic stresses and gravity, {as well as potential artifacts introduced by windowing, smoothing, and projection effects.}

\item The scale-dependent amplitudes of hot atmospheric velocities are derived indirectly from the surface brightness fluctuations using Chandra X-ray observations within the KCWI [OII] emission regions. The velocity amplitudes are $\rm 100 - 200~km~s^{-1}$, comparable to direct Hitomi measurements of ICM velocity dispersion in the Perseus cluster. Due to  Poisson noise, measurements cannot be made on scales accessible by KCWI data. However, we find a consistent and similar trend in amplitude at large spatial scales in Abell 1835 and RXJ0820, suggesting that the cold phase gas condensed from the hot atmospheres.

    \item Using simulations of Gaussian random fields with known underlying power spectra, we explore the impact of a restricted field of view on VSF measurements. We find that when the spatial scale measured exceeds half the width of the observation area, the declining number of pixel pairs can distort the shape of the VSFs, potentially introducing unphysical features. Therefore, characteristics appearing on large scales in VSFs should be interpreted with caution. Given the observational constraints of KCWI and ALMA, we are unable to precisely constrain an energy injection scale leading to turbulence. However, a careful examination reveals that the turnovers in the KCWI [OII] emission VSFs in Abell 1835, RXJ0820, and PKS 0745 correspond closely with the characteristic scales of X-ray cavities present in the system. This suggests that AGN activity may drive the gas motion in the core and could play a significant role in energy injection and transfer.

   \item  The VSFs comparisons between data taken with different instruments must account for differences in spatial and spectral resolution. We find that coarse spatial resolution and smoothing of the data tend to suppress the VSF amplitude, particularly at spatial scales smaller than the PSF and/or the smoothing kernel, which in turn leads to VSF steepening. 
\end{itemize}

\begin{acknowledgments}
We acquired our data at the W. M. Keck Observatory, which is collaboratively managed by the California Institute of Technology, the University of California, and the National Aeronautics and Space Administration, thanks to the substantial support from the W. M. Keck Foundation. We respectfully acknowledge the deep cultural significance of Maunakea for the indigenous Hawaiian community, a site that provides us with the exceptional opportunity to perform our observations. We thank Tom Rose for his valuable discussions regarding ALMA data reduction. B.R.M is grateful for the support from the Natural Sciences and Engineering Research Council of Canada and the Canadian Space Agency Space Science Enhancement Program. A.~L.~C. appreciates the support from the Ingrid and Joseph W. Hibben Endowed Chair at UC San Diego.

\end{acknowledgments}

%

\vspace{5mm}


\software{Astropy \citep{Astropy1,Astropy2,Astropy3},  Python \citep{Python3}, Numpy \citep{Numpy1,Numpy2}, Matplotlib \citep{Matplotlib}, Scipy \citep{SciPy}, 
          }



\appendix

\section{Statistical Uncertainty in Velocity Structure Function Analysis}

\subsection{Window effect on the VSF}
\label{appdenix:window}

The smallest measurable scale in our analysis is constrained by the observational PSF, while the largest scale is limited by the FOV of the data. As the separation increases, the number of pixel pairs decreases, thereby reducing the statistical reliability of the VSF and leading to distortion at larger scales. To mitigate these statistical uncertainties, we include only measurements at scales {where bins exceed 20\% of peak values (with a minimum of 2000 pixel pairs for KCWI [OII] VSFs).} Additionally, any finite-sized dataset has a `window function' due to its boundaries, which can distort the true VSF shape. 
{The scales at which VSFs exhibit turnovers or changes in slope are indicative of the underlying drivers of motion. This is particularly evident when these turnovers correspond to distinct physical features, such as the sizes of X-ray cavities within the system, suggesting a direct relationship between gas motions and AGN feedback mechanisms.} However, if the VSF turnover scale approaches the image extent, it raises questions about their {realiability}. To determine whether these features represent true physical phenomena or artifacts, we used simulated Gaussian random fields to explore the influence of the dataset's limited size on the VSF.

\begin{figure*}
\centering
	\includegraphics[width=17.5cm,height=8.5cm]{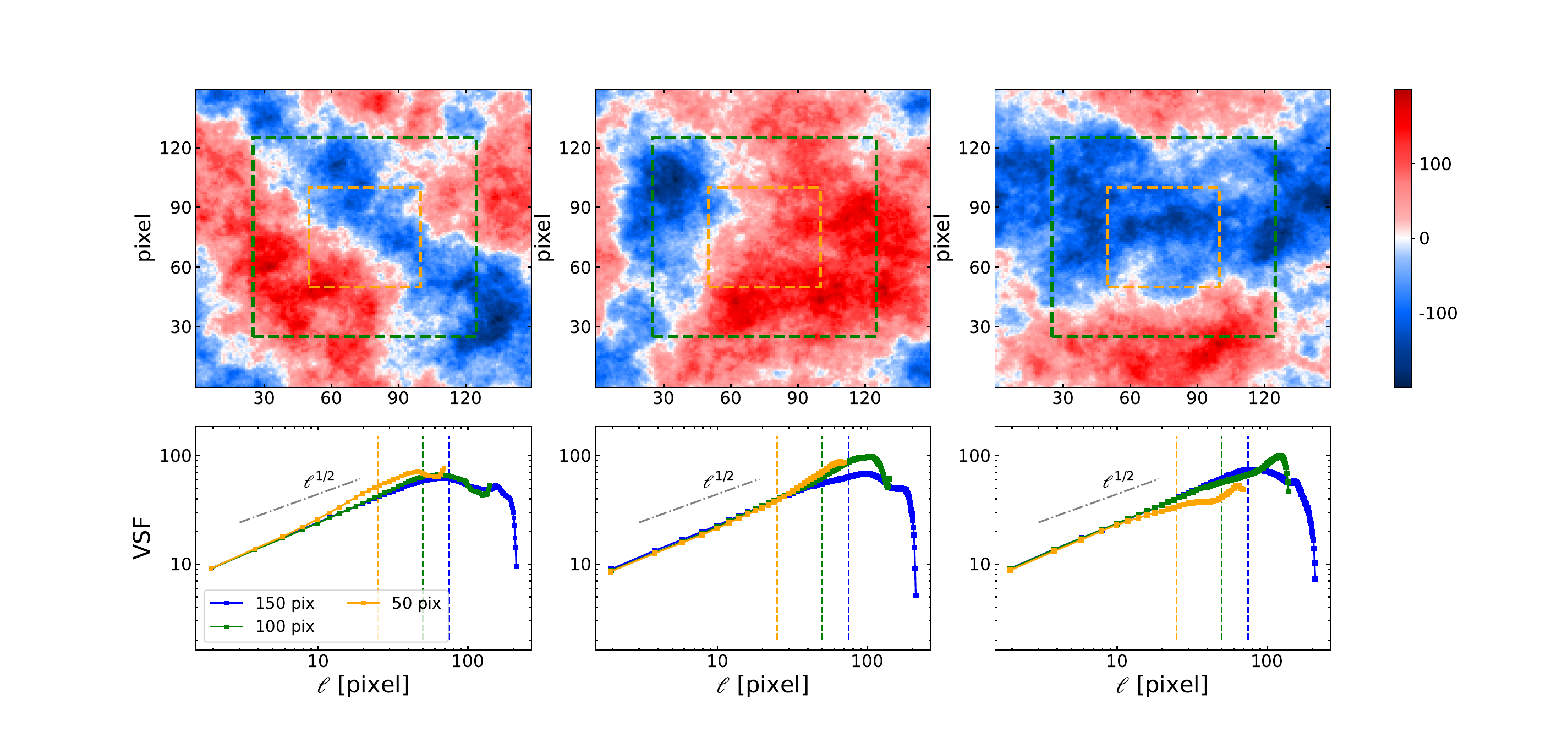}
    \caption{{Window effect} on VSFs. The top panels present a Gaussian random field with a power spectrum of \(\alpha=1/2\), with dimensions of \(150 \times 150\) pixels. The boxes shown on the plot indicate the sub-regions used to examine the effects of finite size. The orange box has a width of 50 pixels, while the green box has a width of 100 pixels. The VSFs calculated from the simulated map and the sub-regions are displayed in the bottom panels. The blue line represents the VSF calculated from the entire image. The VSFs of the inner 50-pixel box, 100-pixel box are indicated by the orange and green lines, respectively. The three vertical lines indicate the half-width of the corresponding windows used for analysis. The grey dash-dotted line represents the expected VSF slope of \(1/2\) for reference.}
    \label{fig:window_effect}
    \end{figure*}

The blue line depicts the VSF derived from the entire map, while the orange and green lines represent the VSF for the \(50\)-pixel and \(100\)-pixel boxes, respectively. A grey dash-dotted reference line with a slope of \(1/2\) indicates the expected VSF slope underlying the Gaussian random field shown above. At smaller scales, the VSFs across all areas closely follow the \(1/2\) power-law behavior, confirming the VSF analysis's ability to accurately recover the underlying two-point distribution. However, turnovers and deviations from the \(1/2\) slope are observed at larger scales near the half-widths of the analysis window, marked by the dashed vertical lines in their respective colors. These distortions, observed in all \(50\) realizations, result from the finite size of the regions from which the VSF is calculated. We conclude that the decrease in pair numbers at larger separations, along with window effects, causes changes in VSF slope and distortions in VSF shape. Therefore, the features in measured VSFs are considered reliable only up to scales no larger than half the image extent.


\subsection{Image smoothing effect on VSF}
\label{appdenix:smoothing}
We compare the VSF computed from ALMA and KCWI observations to study the coupling between the motion of warm and cold gas, as seen in Figure~\ref{fig:alma_keck_vsf_comparison}. The higher spatial resolution of CO ALMA observations allows the VSF measurements at smaller scales than [OII] emission detected by KCWI. The PSF and image smoothing during data reduction introduce correlations between adjacent pixels, that thus affect velocity difference measurements, and change the shape and slope of the VSFs.

To examine the smoothing effect on the VSF measurement, we generated multiple \(50 \times 50\) pixel simulated Gaussian random fields with a power-law power spectrum characterized by an underlying power of 1/2. One realization is depicted in the top row of Figure ~\ref{fig:smooth_effect_on_simulation}. The left panel shows the original, unsmoothed image, while the middle and right panels display the image smoothed by Gaussian kernel of $\sigma=$ 5 pixel width and $\sigma=$ 15 pixel width, respectively. From left to right, the progressive blurring of delicate structures and the apparent reduction in image fluctuations are evident.

The VSFs, computed from the simulated fields smoothed at different levels, are presented in the left panel of the bottom row in Figure.~\ref{fig:smooth_effect_on_simulation}. The VSF of the unsmoothed image (shown as the red line) maintains a slope of approximately 1/2 up to scales of $\sim$20 pixel. 
    For the VSFs of smoothed images, as represented by the green and blue lines, we observe an amplitude suppression relative to the true VSF. The amplitude suppression is consistent with the reduced fluctuations due to smoothing, which decreases the differences between pixels. This impact is dependent on the scales measured and the kernel width. The amplitudes are more heavily suppressed at small scales comparable to or less than the kernel width, leading to a slope steepening, and converging at larger scales. The steepest slope observed in the VSF of the image smoothed by a 15-pixel width kernel (blue line), resulting from the heaviest suppression of the smoothing effect at small scales. 

The scale-dependent impact is clearly presented by the VSF ratio of the smoothed to the unsmoothed images, as shown in the right panel of the bottom row in Figure ~\ref{fig:smooth_effect_on_simulation}. The green line illustrates the mean ratio of 50 VSFs obtained from the unsmoothed images to those from images blurred by a 5-pixel width Gaussian kernel, while the blue line represents the ratio for images blurred by a 15-pixel width Gaussian kernel. The vertical dashed lines indicate the widths of the kernels. The amplitudes are reduced more significantly at scales below the kernel widths, with ratios becoming almost constant at larger scales and approaching 1 as they near the largest pixel separation detectable. The range of scales impacted broadens as the kernel width increases.

This test not only highlights the impact of image smoothing during data reduction but also emphasizes the influence of the instrumental PSF. Notably, the observed VSF provides only an upper limit on the intrinsic slope. As the spatial resolution of telescopes improves, we can get closer to the intrinsic slope. However, even though the measurements on scales below which pixels are strongly correlated are excluded, the resulting distortions still leak to larger scales and cannot be thoroughly avoided. Moreover, the smoothing effect must be considered when comparing the VSFs obtained from different telescopes. To ensure accurate and effective comparison, we applied a box kernel to smooth the ALMA observations, making their beam size match the seeing limits of the KCWI data in Section~\ref{sec: alma vsf}.

\begin{figure*}
\centering
\includegraphics[width=17.5cm,height=10.5cm]{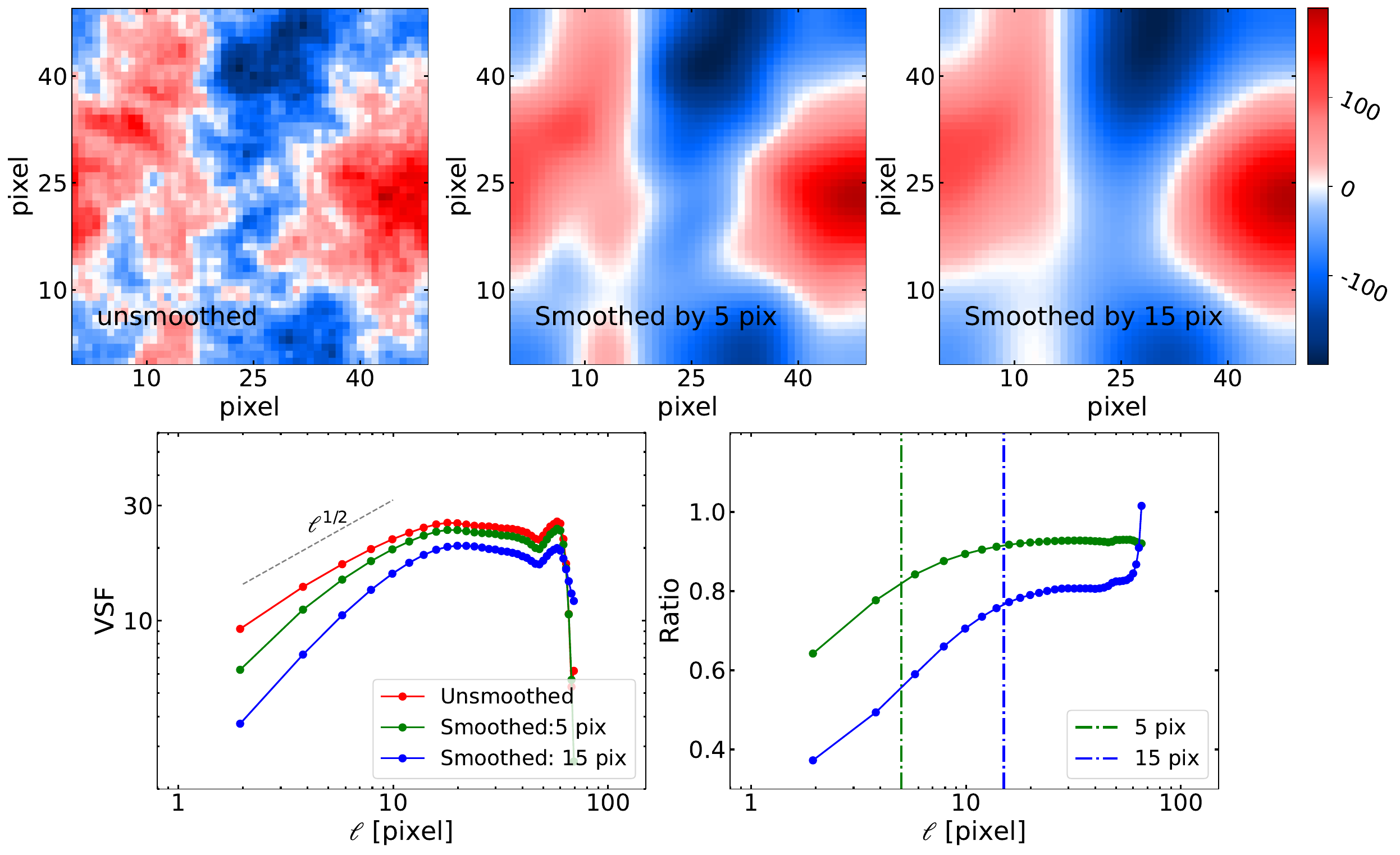}
    \caption{Smoothing effect on VSFs: {Top}: A Gaussian field with a power spectrum characterized by an underlying power of 1/2. To investigate the effect of image smoothing on VSF, the original image (left) is subjected to Gaussian kernel smoothing with a kernel size of 5 pixels (middle) and a kernel size of 15 pixels (right). {Bottom left}: The corresponding VSFs are calculated from simulated Gaussian fields. The red curve represents the unsmoothed image, displaying a slope close to 1/2. The green and blue curves correspond to images smoothed with Gaussian kernels of 5-pixel and 15-pixel widths, respectively, revealing a progressive amplitude suppression, particularly on smaller scales. {Bottom right}: Exploring the impact of various smoothing levels on resulting VSFs. The green line and blue line illustrate the mean ratio of 50 VSFs of the unsmoothed images to those of the images blurred by a 5-pixel width Gaussian kernel and to those of the images blurred by a 15-pixel width Gaussian kernel, respectively.}
    \label{fig:smooth_effect_on_simulation}
\end{figure*}

\bibliography{reference}{}

\begin{thebibliography}{}
\expandafter\ifx\csname natexlab\endcsname\relax\def\natexlab#1{#1}\fi
\providecommand{\url}[1]{\href{#1}{#1}}
\providecommand{\dodoi}[1]{doi:~\href{http://doi.org/#1}{\nolinkurl{#1}}}
\providecommand{\doeprint}[1]{\href{http://ascl.net/#1}{\nolinkurl{http://ascl.net/#1}}}
\providecommand{\doarXiv}[1]{\href{https://arxiv.org/abs/#1}{\nolinkurl{https://arxiv.org/abs/#1}}}

\bibitem[{{Ar{\'e}valo} {et~al.}(2012){Ar{\'e}valo}, {Churazov}, {Zhuravleva}, {Hern{\'a}ndez-Monteagudo}, \& {Revnivtsev}}]{arevalo2012mexican}
{Ar{\'e}valo}, P., {Churazov}, E., {Zhuravleva}, I., {Hern{\'a}ndez-Monteagudo}, C., \& {Revnivtsev}, M. 2012, \mnras, 426, 1793, \dodoi{10.1111/j.1365-2966.2012.21789.x}

\bibitem[{{Astropy Collaboration} {et~al.}(2013){Astropy Collaboration}, {Robitaille}, {Tollerud}, {Greenfield}, {Droettboom}, {Bray}, {Aldcroft}, {Davis}, {Ginsburg}, {Price-Whelan}, {Kerzendorf}, {Conley}, {Crighton}, {Barbary}, {Muna}, {Ferguson}, {Grollier}, {Parikh}, {Nair}, {Unther}, {Deil}, {Woillez}, {Conseil}, {Kramer}, {Turner}, {Singer}, {Fox}, {Weaver}, {Zabalza}, {Edwards}, {Azalee Bostroem}, {Burke}, {Casey}, {Crawford}, {Dencheva}, {Ely}, {Jenness}, {Labrie}, {Lim}, {Pierfederici}, {Pontzen}, {Ptak}, {Refsdal}, {Servillat}, \& {Streicher}}]{Astropy1}
{Astropy Collaboration}, {Robitaille}, T.~P., {Tollerud}, E.~J., {et~al.} 2013, \aap, 558, A33, \dodoi{10.1051/0004-6361/201322068}

\bibitem[{{Astropy Collaboration} {et~al.}(2018){Astropy Collaboration}, {Price-Whelan}, {Sip{\H{o}}cz}, {G{\"u}nther}, {Lim}, {Crawford}, {Conseil}, {Shupe}, {Craig}, {Dencheva}, {Ginsburg}, {VanderPlas}, {Bradley}, {P{\'e}rez-Su{\'a}rez}, {de Val-Borro}, {Aldcroft}, {Cruz}, {Robitaille}, {Tollerud}, {Ardelean}, {Babej}, {Bach}, {Bachetti}, {Bakanov}, {Bamford}, {Barentsen}, {Barmby}, {Baumbach}, {Berry}, {Biscani}, {Boquien}, {Bostroem}, {Bouma}, {Brammer}, {Bray}, {Breytenbach}, {Buddelmeijer}, {Burke}, {Calderone}, {Cano Rodr{\'\i}guez}, {Cara}, {Cardoso}, {Cheedella}, {Copin}, {Corrales}, {Crichton}, {D'Avella}, {Deil}, {Depagne}, {Dietrich}, {Donath}, {Droettboom}, {Earl}, {Erben}, {Fabbro}, {Ferreira}, {Finethy}, {Fox}, {Garrison}, {Gibbons}, {Goldstein}, {Gommers}, {Greco}, {Greenfield}, {Groener}, {Grollier}, {Hagen}, {Hirst}, {Homeier}, {Horton}, {Hosseinzadeh}, {Hu}, {Hunkeler}, {Ivezi{\'c}}, {Jain}, {Jenness}, {Kanarek}, {Kendrew}, {Kern}, {Kerzendorf}, {Khvalko}, {King}, {Kirkby}, {Kulkarni},
  {Kumar}, {Lee}, {Lenz}, {Littlefair}, {Ma}, {Macleod}, {Mastropietro}, {McCully}, {Montagnac}, {Morris}, {Mueller}, {Mumford}, {Muna}, {Murphy}, {Nelson}, {Nguyen}, {Ninan}, {N{\"o}the}, {Ogaz}, {Oh}, {Parejko}, {Parley}, {Pascual}, {Patil}, {Patil}, {Plunkett}, {Prochaska}, {Rastogi}, {Reddy Janga}, {Sabater}, {Sakurikar}, {Seifert}, {Sherbert}, {Sherwood-Taylor}, {Shih}, {Sick}, {Silbiger}, {Singanamalla}, {Singer}, {Sladen}, {Sooley}, {Sornarajah}, {Streicher}, {Teuben}, {Thomas}, {Tremblay}, {Turner}, {Terr{\'o}n}, {van Kerkwijk}, {de la Vega}, {Watkins}, {Weaver}, {Whitmore}, {Woillez}, {Zabalza}, \& {Astropy Contributors}}]{Astropy2}
{Astropy Collaboration}, {Price-Whelan}, A.~M., {Sip{\H{o}}cz}, B.~M., {et~al.} 2018, \aj, 156, 123, \dodoi{10.3847/1538-3881/aabc4f}

\bibitem[{{Astropy Collaboration} {et~al.}(2022){Astropy Collaboration}, {Price-Whelan}, {Lim}, {Earl}, {Starkman}, {Bradley}, {Shupe}, {Patil}, {Corrales}, {Brasseur}, {N{\"o}the}, {Donath}, {Tollerud}, {Morris}, {Ginsburg}, {Vaher}, {Weaver}, {Tocknell}, {Jamieson}, {van Kerkwijk}, {Robitaille}, {Merry}, {Bachetti}, {G{\"u}nther}, {Aldcroft}, {Alvarado-Montes}, {Archibald}, {B{\'o}di}, {Bapat}, {Barentsen}, {Baz{\'a}n}, {Biswas}, {Boquien}, {Burke}, {Cara}, {Cara}, {Conroy}, {Conseil}, {Craig}, {Cross}, {Cruz}, {D'Eugenio}, {Dencheva}, {Devillepoix}, {Dietrich}, {Eigenbrot}, {Erben}, {Ferreira}, {Foreman-Mackey}, {Fox}, {Freij}, {Garg}, {Geda}, {Glattly}, {Gondhalekar}, {Gordon}, {Grant}, {Greenfield}, {Groener}, {Guest}, {Gurovich}, {Handberg}, {Hart}, {Hatfield-Dodds}, {Homeier}, {Hosseinzadeh}, {Jenness}, {Jones}, {Joseph}, {Kalmbach}, {Karamehmetoglu}, {Ka{\l}uszy{\'n}ski}, {Kelley}, {Kern}, {Kerzendorf}, {Koch}, {Kulumani}, {Lee}, {Ly}, {Ma}, {MacBride}, {Maljaars}, {Muna}, {Murphy}, {Norman},
  {O'Steen}, {Oman}, {Pacifici}, {Pascual}, {Pascual-Granado}, {Patil}, {Perren}, {Pickering}, {Rastogi}, {Roulston}, {Ryan}, {Rykoff}, {Sabater}, {Sakurikar}, {Salgado}, {Sanghi}, {Saunders}, {Savchenko}, {Schwardt}, {Seifert-Eckert}, {Shih}, {Jain}, {Shukla}, {Sick}, {Simpson}, {Singanamalla}, {Singer}, {Singhal}, {Sinha}, {Sip{\H{o}}cz}, {Spitler}, {Stansby}, {Streicher}, {{\v{S}}umak}, {Swinbank}, {Taranu}, {Tewary}, {Tremblay}, {de Val-Borro}, {Van Kooten}, {Vasovi{\'c}}, {Verma}, {de Miranda Cardoso}, {Williams}, {Wilson}, {Winkel}, {Wood-Vasey}, {Xue}, {Yoachim}, {Zhang}, {Zonca}, \& {Astropy Project Contributors}}]{Astropy3}
{Astropy Collaboration}, {Price-Whelan}, A.~M., {Lim}, P.~L., {et~al.} 2022, \apj, 935, 167, \dodoi{10.3847/1538-4357/ac7c74}

\bibitem[{{Bambic} {et~al.}(2018){Bambic}, {Morsony}, \& {Reynolds}}]{bambic2018suppression}
{Bambic}, C.~J., {Morsony}, B.~J., \& {Reynolds}, C.~S. 2018, \apj, 857, 84, \dodoi{10.3847/1538-4357/aab558}

\bibitem[{{Bayer-Kim} {et~al.}(2002){Bayer-Kim}, {Crawford}, {Allen}, {Edge}, \& {Fabian}}]{kim2002peculiar}
{Bayer-Kim}, C.~M., {Crawford}, C.~S., {Allen}, S.~W., {Edge}, A.~C., \& {Fabian}, A.~C. 2002, \mnras, 337, 938, \dodoi{10.1046/j.1365-8711.2002.05969.x}

\bibitem[{{B{\^\i}rzan} {et~al.}(2012){B{\^\i}rzan}, {Rafferty}, {Nulsen}, {McNamara}, {R{\"o}ttgering}, {Wise}, \& {Mittal}}]{birzan2012duty}
{B{\^\i}rzan}, L., {Rafferty}, D.~A., {Nulsen}, P.~E.~J., {et~al.} 2012, \mnras, 427, 3468, \dodoi{10.1111/j.1365-2966.2012.22083.x}

\bibitem[{{Cappellari} \& {Emsellem}(2004)}]{PPXF}
{Cappellari}, M., \& {Emsellem}, E. 2004, \pasp, 116, 138, \dodoi{10.1086/381875}

\bibitem[{{Cavagnolo} {et~al.}(2008){Cavagnolo}, {Donahue}, {Voit}, \& {Sun}}]{cavagnolo2008entropy}
{Cavagnolo}, K.~W., {Donahue}, M., {Voit}, G.~M., \& {Sun}, M. 2008, \apjl, 683, L107, \dodoi{10.1086/591665}

\bibitem[{{Chen} {et~al.}(2023){Chen}, {Chen}, {Rauch}, {Qu}, {Johnson}, {Li}, {Schaye}, {Rudie}, {Zahedy}, {Boettcher}, {Cooksey}, \& {Cantalupo}}]{chen2023empirical}
{Chen}, M.~C., {Chen}, H.-W., {Rauch}, M., {et~al.} 2023, \mnras, 518, 2354, \dodoi{10.1093/mnras/stac3193}

\bibitem[{{Churazov} {et~al.}(2002){Churazov}, {Sunyaev}, {Forman}, \& {B{\"o}hringer}}]{churazov2002cooling}
{Churazov}, E., {Sunyaev}, R., {Forman}, W., \& {B{\"o}hringer}, H. 2002, \mnras, 332, 729, \dodoi{10.1046/j.1365-8711.2002.05332.x}

\bibitem[{{Churazov} {et~al.}(2012){Churazov}, {Vikhlinin}, {Zhuravleva}, {Schekochihin}, {Parrish}, {Sunyaev}, {Forman}, {B{\"o}hringer}, \& {Randall}}]{churazov2012x}
{Churazov}, E., {Vikhlinin}, A., {Zhuravleva}, I., {et~al.} 2012, \mnras, 421, 1123, \dodoi{10.1111/j.1365-2966.2011.20372.x}

\bibitem[{{Ciotti} \& {Ostriker}(2001)}]{ciotti2001cooling}
{Ciotti}, L., \& {Ostriker}, J.~P. 2001, \apj, 551, 131, \dodoi{10.1086/320053}

\bibitem[{{Clarke} {et~al.}(2009){Clarke}, {Blanton}, {Sarazin}, {Anderson}, {Gopal-Krishna}, {Douglass}, \& {Kassim}}]{clarke2009tracing}
{Clarke}, T.~E., {Blanton}, E.~L., {Sarazin}, C.~L., {et~al.} 2009, \apj, 697, 1481, \dodoi{10.1088/0004-637X/697/2/1481}

\bibitem[{{Cowie} \& {Binney}(1977)}]{1977ApJ...215..723C}
{Cowie}, L.~L., \& {Binney}, J. 1977, \apj, 215, 723, \dodoi{10.1086/155406}

\bibitem[{{Crawford} {et~al.}(1999){Crawford}, {Allen}, {Ebeling}, {Edge}, \& {Fabian}}]{crawford1999rosat}
{Crawford}, C.~S., {Allen}, S.~W., {Ebeling}, H., {Edge}, A.~C., \& {Fabian}, A.~C. 1999, \mnras, 306, 857, \dodoi{10.1046/j.1365-8711.1999.02583.x}

\bibitem[{{Edge}(2001)}]{edge2001detection}
{Edge}, A.~C. 2001, \mnras, 328, 762, \dodoi{10.1046/j.1365-8711.2001.04802.x}

\bibitem[{{Fabian}(2012)}]{2012ARA&A..50..455F}
{Fabian}, A.~C. 2012, \araa, 50, 455, \dodoi{10.1146/annurev-astro-081811-125521}

\bibitem[{{Fabian} {et~al.}(2006){Fabian}, {Sanders}, {Taylor}, {Allen}, {Crawford}, {Johnstone}, \& {Iwasawa}}]{fabian2006very}
{Fabian}, A.~C., {Sanders}, J.~S., {Taylor}, G.~B., {et~al.} 2006, \mnras, 366, 417, \dodoi{10.1111/j.1365-2966.2005.09896.x}

\bibitem[{{Fournier} {et~al.}(2025){Fournier}, {Grete}, {Bruggen}, {Glines}, {O'Shea}, {Prasad}, \& {Wibking}}]{fournier2025vsf}
{Fournier}, M., {Grete}, P., {Bruggen}, M., {et~al.} 2025, {Submitted to A\&A, `XMAGNET: Velocity structure functions in AGN-driven turbulence of the multiphase intracluster medium' }

\bibitem[{{Ganguly} {et~al.}(2023){Ganguly}, {Li}, {Olivares}, {Su}, {Combes}, {Prakash}, {Hamer}, {Guillard}, \& {Ha}}]{ganguly2023nature}
{Ganguly}, S., {Li}, Y., {Olivares}, V., {et~al.} 2023, Frontiers in Astronomy and Space Sciences, 10, 1138613, \dodoi{10.3389/fspas.2023.1138613}

\bibitem[{{Gaspari} {et~al.}(2014){Gaspari}, {Churazov}, {Nagai}, {Lau}, \& {Zhuravleva}}]{gaspari2014relation}
{Gaspari}, M., {Churazov}, E., {Nagai}, D., {Lau}, E.~T., \& {Zhuravleva}, I. 2014, \aap, 569, A67, \dodoi{10.1051/0004-6361/201424043}

\bibitem[{{Gaspari} {et~al.}(2017){Gaspari}, {Temi}, \& {Brighenti}}]{gaspari2017raining}
{Gaspari}, M., {Temi}, P., \& {Brighenti}, F. 2017, \mnras, 466, 677, \dodoi{10.1093/mnras/stw3108}

\bibitem[{{Gaspari} {et~al.}(2018){Gaspari}, {McDonald}, {Hamer}, {Brighenti}, {Temi}, {Gendron-Marsolais}, {Hlavacek-Larrondo}, {Edge}, {Werner}, {Tozzi}, {Sun}, {Stone}, {Tremblay}, {Hogan}, {Eckert}, {Ettori}, {Yu}, {Biffi}, \& {Planelles}}]{gaspari2018shaken}
{Gaspari}, M., {McDonald}, M., {Hamer}, S.~L., {et~al.} 2018, \apj, 854, 167, \dodoi{10.3847/1538-4357/aaaa1b}

\bibitem[{{Gingras} {et~al.}(2024){Gingras}, {Coil}, {McNamara}, {Perrotta}, {Brighenti}, {Russell}, {Li}, {Oh}, \& {Ning}}]{gingras2024complex}
{Gingras}, M.-J., {Coil}, A.~L., {McNamara}, B.~R., {et~al.} 2024, arXiv e-prints, arXiv:2404.02212, \dodoi{10.48550/arXiv.2404.02212}

\bibitem[{{Gonz{\'a}lez Delgado} {et~al.}(2005){Gonz{\'a}lez Delgado}, {Cervi{\~n}o}, {Martins}, {Leitherer}, \& {Hauschildt}}]{SSP2005}
{Gonz{\'a}lez Delgado}, R.~M., {Cervi{\~n}o}, M., {Martins}, L.~P., {Leitherer}, C., \& {Hauschildt}, P.~H. 2005, \mnras, 357, 945, \dodoi{10.1111/j.1365-2966.2005.08692.x}

\bibitem[{{Guo} \& {Oh}(2008)}]{guo2008feedback}
{Guo}, F., \& {Oh}, S.~P. 2008, \mnras, 384, 251, \dodoi{10.1111/j.1365-2966.2007.12692.x}

\bibitem[{{Harris} {et~al.}(2020){Harris}, {Millman}, {van der Walt}, {Gommers}, {Virtanen}, {Cournapeau}, {Wieser}, {Taylor}, {Berg}, {Smith}, {Kern}, {Picus}, {Hoyer}, {van Kerkwijk}, {Brett}, {Haldane}, {del R{\'\i}o}, {Wiebe}, {Peterson}, {G{\'e}rard-Marchant}, {Sheppard}, {Reddy}, {Weckesser}, {Abbasi}, {Gohlke}, \& {Oliphant}}]{Numpy2}
{Harris}, C.~R., {Millman}, K.~J., {van der Walt}, S.~J., {et~al.} 2020, \nat, 585, 357, \dodoi{10.1038/s41586-020-2649-2}

\bibitem[{{Heckman} {et~al.}(1989){Heckman}, {Baum}, {van Breugel}, \& {McCarthy}}]{1989ApJ...338...48H}
{Heckman}, T.~M., {Baum}, S.~A., {van Breugel}, W.~J.~M., \& {McCarthy}, P. 1989, \apj, 338, 48, \dodoi{10.1086/167181}

\bibitem[{{Heinz} {et~al.}(1998){Heinz}, {Reynolds}, \& {Begelman}}]{heinz1998x}
{Heinz}, S., {Reynolds}, C.~S., \& {Begelman}, M.~C. 1998, \apj, 501, 126, \dodoi{10.1086/305807}

\bibitem[{{Hillel} \& {Soker}(2016)}]{hillel2016heating}
{Hillel}, S., \& {Soker}, N. 2016, \mnras, 455, 2139, \dodoi{10.1093/mnras/stv2483}

\bibitem[{{Hillel} \& {Soker}(2017)}]{hillel2017hitomi}
---. 2017, \mnras, 466, L39, \dodoi{10.1093/mnrasl/slw231}

\bibitem[{{Hillel} \& {Soker}(2020)}]{hillel2020kinematics}
---. 2020, \apj, 896, 104, \dodoi{10.3847/1538-4357/ab9109}

\bibitem[{{Hitomi Collaboration} {et~al.}(2016){Hitomi Collaboration}, {Aharonian}, {Akamatsu}, {Akimoto}, {Allen}, {Anabuki}, {Angelini}, {Arnaud}, {Audard}, {Awaki}, {Axelsson}, {Bamba}, {Bautz}, {Blandford}, {Brenneman}, {Brown}, {Bulbul}, {Cackett}, {Chernyakova}, {Chiao}, {Coppi}, {Costantini}, {de Plaa}, {den Herder}, {Done}, {Dotani}, {Ebisawa}, {Eckart}, {Enoto}, {Ezoe}, {Fabian}, {Ferrigno}, {Foster}, {Fujimoto}, {Fukazawa}, {Furuzawa}, {Galeazzi}, {Gallo}, {Gandhi}, {Giustini}, {Goldwurm}, {Gu}, {Guainazzi}, {Haba}, {Hagino}, {Hamaguchi}, {Harrus}, {Hatsukade}, {Hayashi}, {Hayashi}, {Hayashida}, {Hiraga}, {Hornschemeier}, {Hoshino}, {Hughes}, {Iizuka}, {Inoue}, {Inoue}, {Ishibashi}, {Ishida}, {Ishikawa}, {Ishisaki}, {Itoh}, {Iyomoto}, {Kaastra}, {Kallman}, {Kamae}, {Kara}, {Kataoka}, {Katsuda}, {Katsuta}, {Kawaharada}, {Kawai}, {Kelley}, {Khangulyan}, {Kilbourne}, {King}, {Kitaguchi}, {Kitamoto}, {Kitayama}, {Kohmura}, {Kokubun}, {Koyama}, {Koyama}, {Kretschmar}, {Krimm}, {Kubota}, {Kunieda},
  {Laurent}, {Lebrun}, {Lee}, {Leutenegger}, {Limousin}, {Loewenstein}, {Long}, {Lumb}, {Madejski}, {Maeda}, {Maier}, {Makishima}, {Markevitch}, {Matsumoto}, {Matsushita}, {McCammon}, {McNamara}, {Mehdipour}, {Miller}, {Miller}, {Mineshige}, {Mitsuda}, {Mitsuishi}, {Miyazawa}, {Mizuno}, {Mori}, {Mori}, {Moseley}, {Mukai}, {Murakami}, {Murakami}, {Mushotzky}, {Nagino}, {Nakagawa}, {Nakajima}, {Nakamori}, {Nakano}, {Nakashima}, {Nakazawa}, {Nobukawa}, {Noda}, {Nomachi}, {O'Dell}, {Odaka}, {Ohashi}, {Ohno}, {Okajima}, {Ota}, {Ozaki}, {Paerels}, {Paltani}, {Parmar}, {Petre}, {Pinto}, {Pohl}, {Porter}, {Pottschmidt}, {Ramsey}, {Reynolds}, {Russell}, {Safi-Harb}, {Saito}, {Sakai}, {Sameshima}, {Sato}, {Sato}, {Sato}, {Sawada}, {Schartel}, {Serlemitsos}, {Seta}, {Shidatsu}, {Simionescu}, {Smith}, {Soong}, {Stawarz}, {Sugawara}, {Sugita}, {Szymkowiak}, {Tajima}, {Takahashi}, {Takahashi}, {Takeda}, {Takei}, {Tamagawa}, {Tamura}, {Tamura}, {Tanaka}, {Tanaka}, {Tanaka}, {Tashiro}, {Tawara}, {Terada}, {Terashima},
  {Tombesi}, {Tomida}, {Tsuboi}, {Tsujimoto}, {Tsunemi}, {Tsuru}, {Uchida}, {Uchiyama}, {Uchiyama}, {Ueda}, {Ueda}, {Ueno}, {Uno}, {Urry}, {Ursino}, {de Vries}, {Watanabe}, {Werner}, {Wik}, {Wilkins}, {Williams}, {Yamada}, {Yamaguchi}, {Yamaoka}, {Yamasaki}, {Yamauchi}, {Yamauchi}, {Yaqoob}, {Yatsu}, {Yonetoku}, {Yoshida}, {Yuasa}, {Zhuravleva}, \& {Zoghbi}}]{hitomi2016quiescent}
{Hitomi Collaboration}, {Aharonian}, F., {Akamatsu}, H., {et~al.} 2016, \nat, 535, 117, \dodoi{10.1038/nature18627}

\bibitem[{{Hu} {et~al.}(2022){Hu}, {Qiu}, {Gendron-Marsolais}, {Bogdanovi{\'c}}, {Hlavacek-Larrondo}, {Ho}, {Inayoshi}, \& {McNamara}}]{2022ApJ...929L..30H}
{Hu}, H., {Qiu}, Y., {Gendron-Marsolais}, M.-L., {et~al.} 2022, \apjl, 929, L30, \dodoi{10.3847/2041-8213/ac6601}

\bibitem[{{Hunter}(2007)}]{Matplotlib}
{Hunter}, J.~D. 2007, Computing in Science and Engineering, 9, 90, \dodoi{10.1109/MCSE.2007.55}

\bibitem[{{Jaffe} {et~al.}(2005){Jaffe}, {Bremer}, \& {Baker}}]{jaffe2005h}
{Jaffe}, W., {Bremer}, M.~N., \& {Baker}, K. 2005, \mnras, 360, 748, \dodoi{10.1111/j.1365-2966.2005.09073.x}

\bibitem[{{Li} {et~al.}(2020){Li}, {Gendron-Marsolais}, {Zhuravleva}, {Xu}, {Simionescu}, {Tremblay}, {Lochhaas}, {Bryan}, {Quataert}, {Murray}, {Boselli}, {Hlavacek-Larrondo}, {Zheng}, {Fossati}, {Li}, {Emsellem}, {Sarzi}, {Arzamasskiy}, \& {Vishniac}}]{2020ApJ...889L...1L}
{Li}, Y., {Gendron-Marsolais}, M.-L., {Zhuravleva}, I., {et~al.} 2020, \apjl, 889, L1, \dodoi{10.3847/2041-8213/ab65c7}

\bibitem[{{Loewenstein} {et~al.}(1991){Loewenstein}, {Zweibel}, \& {Begelman}}]{1991ApJ...377..392L}
{Loewenstein}, M., {Zweibel}, E.~G., \& {Begelman}, M.~C. 1991, \apj, 377, 392, \dodoi{10.1086/170369}

\bibitem[{{Markevitch} \& {Vikhlinin}(2007)}]{2007PhR...443....1M}
{Markevitch}, M., \& {Vikhlinin}, A. 2007, \physrep, 443, 1, \dodoi{10.1016/j.physrep.2007.01.001}

\bibitem[{{Markwardt}(2009)}]{MPFIT}
{Markwardt}, C.~B. 2009, in Astronomical Society of the Pacific Conference Series, Vol. 411, Astronomical Data Analysis Software and Systems XVIII, ed. D.~A. {Bohlender}, D.~{Durand}, \& P.~{Dowler}, 251.
\newblock \doarXiv{0902.2850}

\bibitem[{{McCourt} {et~al.}(2012){McCourt}, {Sharma}, {Quataert}, \& {Parrish}}]{mccourt2012thermal}
{McCourt}, M., {Sharma}, P., {Quataert}, E., \& {Parrish}, I.~J. 2012, \mnras, 419, 3319, \dodoi{10.1111/j.1365-2966.2011.19972.x}

\bibitem[{{McDonald} {et~al.}(2011){McDonald}, {Veilleux}, \& {Mushotzky}}]{mcdonald2011effect}
{McDonald}, M., {Veilleux}, S., \& {Mushotzky}, R. 2011, \apj, 731, 33, \dodoi{10.1088/0004-637X/731/1/33}

\bibitem[{{McLaughlin} \& {Bell}(1998)}]{mclaughlin1998electron}
{McLaughlin}, B.~M., \& {Bell}, K.~L. 1998, Journal of Physics B Atomic Molecular Physics, 31, 4317, \dodoi{10.1088/0953-4075/31/19/017}

\bibitem[{{McNamara} {et~al.}(2016){McNamara}, {Russell}, {Nulsen}, {Hogan}, {Fabian}, {Pulido}, \& {Edge}}]{mcnamara2016mechanism}
{McNamara}, B.~R., {Russell}, H.~R., {Nulsen}, P.~E.~J., {et~al.} 2016, \apj, 830, 79, \dodoi{10.3847/0004-637X/830/2/79}

\bibitem[{{McNamara} {et~al.}(2006){McNamara}, {Rafferty}, {B{\^\i}rzan}, {Steiner}, {Wise}, {Nulsen}, {Carilli}, {Ryan}, \& {Sharma}}]{mcnamara2006starburst}
{McNamara}, B.~R., {Rafferty}, D.~A., {B{\^\i}rzan}, L., {et~al.} 2006, \apj, 648, 164, \dodoi{10.1086/505859}

\bibitem[{{McNamara} {et~al.}(2014){McNamara}, {Russell}, {Nulsen}, {Edge}, {Murray}, {Main}, {Vantyghem}, {Combes}, {Fabian}, {Salome}, {Kirkpatrick}, {Baum}, {Bregman}, {Donahue}, {Egami}, {Hamer}, {O'Dea}, {Oonk}, {Tremblay}, \& {Voit}}]{mcnamara20141010}
{McNamara}, B.~R., {Russell}, H.~R., {Nulsen}, P.~E.~J., {et~al.} 2014, \apj, 785, 44, \dodoi{10.1088/0004-637X/785/1/44}

\bibitem[{{Mohapatra} {et~al.}(2022){Mohapatra}, {Jetti}, {Sharma}, \& {Federrath}}]{mohapatra2022velocity}
{Mohapatra}, R., {Jetti}, M., {Sharma}, P., \& {Federrath}, C. 2022, \mnras, 510, 2327, \dodoi{10.1093/mnras/stab3429}

\bibitem[{{Morrissey} {et~al.}(2018){Morrissey}, {Matuszewski}, {Martin}, {Neill}, {Epps}, {Fucik}, {Weber}, {Darvish}, {Adkins}, {Allen}, {Bartos}, {Belicki}, {Cabak}, {Callahan}, {Cowley}, {Crabill}, {Deich}, {Delecroix}, {Doppman}, {Hilyard}, {James}, {Kaye}, {Kokorowski}, {Kwok}, {Lanclos}, {Milner}, {Moore}, {O'Sullivan}, {Parihar}, {Park}, {Phillips}, {Rizzi}, {Rockosi}, {Rodriguez}, {Salaun}, {Seaman}, {Sheikh}, {Weiss}, \& {Zarzaca}}]{morrissey2018keck}
{Morrissey}, P., {Matuszewski}, M., {Martin}, D.~C., {et~al.} 2018, \apj, 864, 93, \dodoi{10.3847/1538-4357/aad597}

\bibitem[{{Nulsen} \& {Fabian}(2000)}]{nulsen2000fuelling}
{Nulsen}, P.~E.~J., \& {Fabian}, A.~C. 2000, \mnras, 311, 346, \dodoi{10.1046/j.1365-8711.2000.03038.x}

\bibitem[{{Nulsen} \& {McNamara}(2013)}]{2013AN....334..386N}
{Nulsen}, P.~E.~J., \& {McNamara}, B.~R. 2013, Astronomische Nachrichten, 334, 386, \dodoi{10.1002/asna.201211863}

\bibitem[{{O'Dell} \& {Castaneda}(1987)}]{1987ApJ...317..686O}
{O'Dell}, C.~R., \& {Castaneda}, H.~O. 1987, \apj, 317, 686, \dodoi{10.1086/165314}

\bibitem[{{Olivares} {et~al.}(2019){Olivares}, {Salome}, {Combes}, {Hamer}, {Guillard}, {Lehnert}, {Polles}, {Beckmann}, {Dubois}, {Donahue}, {Edge}, {Fabian}, {McNamara}, {Rose}, {Russell}, {Tremblay}, {Vantyghem}, {Canning}, {Ferland}, {Godard}, {Peirani}, \& {Pineau des Forets}}]{olivares2019ubiquitous}
{Olivares}, V., {Salome}, P., {Combes}, F., {et~al.} 2019, \aap, 631, A22, \dodoi{10.1051/0004-6361/201935350}

\bibitem[{{Omma} {et~al.}(2004){Omma}, {Binney}, {Bryan}, \& {Slyz}}]{omma2004heating}
{Omma}, H., {Binney}, J., {Bryan}, G., \& {Slyz}, A. 2004, \mnras, 348, 1105, \dodoi{10.1111/j.1365-2966.2004.07382.x}

\bibitem[{{Peterson} \& {Fabian}(2006)}]{peterson2006x}
{Peterson}, J.~R., \& {Fabian}, A.~C. 2006, \physrep, 427, 1, \dodoi{10.1016/j.physrep.2005.12.007}

\bibitem[{{Pfrommer}(2013)}]{pfrommer2013toward}
{Pfrommer}, C. 2013, \apj, 779, 10, \dodoi{10.1088/0004-637X/779/1/10}

\bibitem[{{Pradhan} {et~al.}(2006){Pradhan}, {Montenegro}, {Nahar}, \& {Eissner}}]{pradhan2006oii}
{Pradhan}, A.~K., {Montenegro}, M., {Nahar}, S.~N., \& {Eissner}, W. 2006, \mnras, 366, L6, \dodoi{10.1111/j.1745-3933.2005.00119.x}

\bibitem[{{Pulido} {et~al.}(2018){Pulido}, {McNamara}, {Edge}, {Hogan}, {Vantyghem}, {Russell}, {Nulsen}, {Babyk}, \& {Salom{\'e}}}]{pulido2018origin}
{Pulido}, F.~A., {McNamara}, B.~R., {Edge}, A.~C., {et~al.} 2018, \apj, 853, 177, \dodoi{10.3847/1538-4357/aaa54b}

\bibitem[{{Qiu} {et~al.}(2020){Qiu}, {Bogdanovi{\'c}}, {Li}, {McDonald}, \& {McNamara}}]{qiu2020formation}
{Qiu}, Y., {Bogdanovi{\'c}}, T., {Li}, Y., {McDonald}, M., \& {McNamara}, B.~R. 2020, Nature Astronomy, 4, 900, \dodoi{10.1038/s41550-020-1090-7}

\bibitem[{{Rafferty} {et~al.}(2008){Rafferty}, {McNamara}, \& {Nulsen}}]{rafferty2008regulation}
{Rafferty}, D.~A., {McNamara}, B.~R., \& {Nulsen}, P.~E.~J. 2008, \apj, 687, 899, \dodoi{10.1086/591240}

\bibitem[{{Rafferty} {et~al.}(2006){Rafferty}, {McNamara}, {Nulsen}, \& {Wise}}]{rafferty2006feedback}
{Rafferty}, D.~A., {McNamara}, B.~R., {Nulsen}, P.~E.~J., \& {Wise}, M.~W. 2006, \apj, 652, 216, \dodoi{10.1086/507672}

\bibitem[{{Randall} {et~al.}(2011){Randall}, {Forman}, {Giacintucci}, {Nulsen}, {Sun}, {Jones}, {Churazov}, {David}, {Kraft}, {Donahue}, {Blanton}, {Simionescu}, \& {Werner}}]{randall2010shocks}
{Randall}, S.~W., {Forman}, W.~R., {Giacintucci}, S., {et~al.} 2011, \apj, 726, 86, \dodoi{10.1088/0004-637X/726/2/86}

\bibitem[{{Russell} {et~al.}(2016){Russell}, {McNamara}, {Fabian}, {Nulsen}, {Edge}, {Combes}, {Murray}, {Parrish}, {Salom{\'e}}, {Sanders}, {Baum}, {Donahue}, {Main}, {O'Connell}, {O'Dea}, {Oonk}, {Tremblay}, {Vantyghem}, \& {Voit}}]{russell2016alma}
{Russell}, H.~R., {McNamara}, B.~R., {Fabian}, A.~C., {et~al.} 2016, \mnras, 458, 3134, \dodoi{10.1093/mnras/stw409}

\bibitem[{{Russell} {et~al.}(2017){Russell}, {McNamara}, {Fabian}, {Nulsen}, {Combes}, {Edge}, {Hogan}, {McDonald}, {Salom{\'e}}, {Tremblay}, \& {Vantyghem}}]{russell2017close}
---. 2017, \mnras, 472, 4024, \dodoi{10.1093/mnras/stx2255}

\bibitem[{{Russell} {et~al.}(2019){Russell}, {McNamara}, {Fabian}, {Nulsen}, {Combes}, {Edge}, {Madar}, {Olivares}, {Salom{\'e}}, \& {Vantyghem}}]{russell2019driving}
---. 2019, \mnras, 490, 3025, \dodoi{10.1093/mnras/stz2719}

\bibitem[{{Salom{\'e}} \& {Combes}(2003)}]{salome2003cold}
{Salom{\'e}}, P., \& {Combes}, F. 2003, \aap, 412, 657, \dodoi{10.1051/0004-6361:20031438}

\bibitem[{{Salom{\'e}} {et~al.}(2008){Salom{\'e}}, {Combes}, {Revaz}, {Edge}, {Hatch}, {Fabian}, \& {Johnstone}}]{salome2008cold}
{Salom{\'e}}, P., {Combes}, F., {Revaz}, Y., {et~al.} 2008, \aap, 484, 317, \dodoi{10.1051/0004-6361:200809493}

\bibitem[{{Sanders} \& {Fabian}(2007)}]{sanders2007deeper}
{Sanders}, J.~S., \& {Fabian}, A.~C. 2007, \mnras, 381, 1381, \dodoi{10.1111/j.1365-2966.2007.12347.x}

\bibitem[{{Sanders} {et~al.}(2014){Sanders}, {Fabian}, {Hlavacek-Larrondo}, {Russell}, {Taylor}, {Hofmann}, {Tremblay}, \& {Walker}}]{sanders2014feedback}
{Sanders}, J.~S., {Fabian}, A.~C., {Hlavacek-Larrondo}, J., {et~al.} 2014, \mnras, 444, 1497, \dodoi{10.1093/mnras/stu1543}

\bibitem[{{Sternberg} \& {Soker}(2009)}]{sternberg2009sound}
{Sternberg}, A., \& {Soker}, N. 2009, \mnras, 395, 228, \dodoi{10.1111/j.1365-2966.2009.14566.x}

\bibitem[{{van der Walt} {et~al.}(2011){van der Walt}, {Colbert}, \& {Varoquaux}}]{Numpy1}
{van der Walt}, S., {Colbert}, S.~C., \& {Varoquaux}, G. 2011, Computing in Science and Engineering, 13, 22, \dodoi{10.1109/MCSE.2011.37}

\bibitem[{Van~Rossum \& Drake(2009)}]{Python3}
Van~Rossum, G., \& Drake, F.~L. 2009, Python 3 Reference Manual (Scotts Valley, CA: CreateSpace)

\bibitem[{{Vantyghem} {et~al.}(2016){Vantyghem}, {McNamara}, {Russell}, {Hogan}, {Edge}, {Nulsen}, {Fabian}, {Combes}, {Salom{\'e}}, {Baum}, {Donahue}, {Main}, {Murray}, {O'Connell}, {O'Dea}, {Oonk}, {Parrish}, {Sanders}, {Tremblay}, \& {Voit}}]{vantyghem2016molecular}
{Vantyghem}, A.~N., {McNamara}, B.~R., {Russell}, H.~R., {et~al.} 2016, \apj, 832, 148, \dodoi{10.3847/0004-637X/832/2/148}

\bibitem[{{Vantyghem} {et~al.}(2019){Vantyghem}, {McNamara}, {Russell}, {Edge}, {Nulsen}, {Combes}, {Fabian}, {McDonald}, \& {Salom{\'e}}}]{vantyghem2019enormous}
---. 2019, \apj, 870, 57, \dodoi{10.3847/1538-4357/aaf1b4}

\bibitem[{{Vikhlinin} {et~al.}(2005){Vikhlinin}, {Markevitch}, {Murray}, {Jones}, {Forman}, \& {Van Speybroeck}}]{vikhlinin2005chandra}
{Vikhlinin}, A., {Markevitch}, M., {Murray}, S.~S., {et~al.} 2005, \apj, 628, 655, \dodoi{10.1086/431142}

\bibitem[{{Virtanen} {et~al.}(2020){Virtanen}, {Gommers}, {Oliphant}, {Haberland}, {Reddy}, {Cournapeau}, {Burovski}, {Peterson}, {Weckesser}, {Bright}, {van der Walt}, {Brett}, {Wilson}, {Millman}, {Mayorov}, {Nelson}, {Jones}, {Kern}, {Larson}, {Carey}, {Polat}, {Feng}, {Moore}, {VanderPlas}, {Laxalde}, {Perktold}, {Cimrman}, {Henriksen}, {Quintero}, {Harris}, {Archibald}, {Ribeiro}, {Pedregosa}, {van Mulbregt}, \& {SciPy 1. 0 Contributors}}]{SciPy}
{Virtanen}, P., {Gommers}, R., {Oliphant}, T.~E., {et~al.} 2020, Nature Methods, 17, 261, \dodoi{10.1038/s41592-019-0686-2}

\bibitem[{{Voit}(2018)}]{voit2018role}
{Voit}, G.~M. 2018, \apj, 868, 102, \dodoi{10.3847/1538-4357/aae8e2}

\bibitem[{{Voit} \& {Donahue}(2005)}]{voit2005observationally}
{Voit}, G.~M., \& {Donahue}, M. 2005, \apj, 634, 955, \dodoi{10.1086/497063}

\bibitem[{{Wang} {et~al.}(2021){Wang}, {Ruszkowski}, {Pfrommer}, {Oh}, \& {Yang}}]{wang2021non}
{Wang}, C., {Ruszkowski}, M., {Pfrommer}, C., {Oh}, S.~P., \& {Yang}, H. Y.~K. 2021, \mnras, 504, 898, \dodoi{10.1093/mnras/stab966}

\bibitem[{{Wilman} {et~al.}(2006){Wilman}, {Edge}, \& {Swinbank}}]{wilman2006integral}
{Wilman}, R.~J., {Edge}, A.~C., \& {Swinbank}, A.~M. 2006, \mnras, 371, 93, \dodoi{10.1111/j.1365-2966.2006.10658.x}

\bibitem[{{Xu}(2020)}]{xu2020projected}
{Xu}, S. 2020, \mnras, 492, 1044, \dodoi{10.1093/mnras/stz3092}

\bibitem[{{Zhang} {et~al.}(2022){Zhang}, {Zhuravleva}, {Gendron-Marsolais}, {Churazov}, {Schekochihin}, \& {Forman}}]{zhang2022bubble}
{Zhang}, C., {Zhuravleva}, I., {Gendron-Marsolais}, M.-L., {et~al.} 2022, \mnras, 517, 616, \dodoi{10.1093/mnras/stac2282}

\bibitem[{{Zhuravleva} {et~al.}(2018){Zhuravleva}, {Allen}, {Mantz}, \& {Werner}}]{zhuravleva2018gas}
{Zhuravleva}, I., {Allen}, S.~W., {Mantz}, A., \& {Werner}, N. 2018, \apj, 865, 53, \dodoi{10.3847/1538-4357/aadae3}

\bibitem[{{Zhuravleva} {et~al.}(2014{\natexlab{a}}){Zhuravleva}, {Churazov}, {Schekochihin}, {Allen}, {Ar{\'e}valo}, {Fabian}, {Forman}, {Sanders}, {Simionescu}, {Sunyaev}, {Vikhlinin}, \& {Werner}}]{zhuravleva2014turbulent}
{Zhuravleva}, I., {Churazov}, E., {Schekochihin}, A.~A., {et~al.} 2014{\natexlab{a}}, \nat, 515, 85, \dodoi{10.1038/nature13830}

\bibitem[{{Zhuravleva} {et~al.}(2014{\natexlab{b}}){Zhuravleva}, {Churazov}, {Schekochihin}, {Lau}, {Nagai}, {Gaspari}, {Allen}, {Nelson}, \& {Parrish}}]{zhuravleva2014relation}
{Zhuravleva}, I., {Churazov}, E.~M., {Schekochihin}, A.~A., {et~al.} 2014{\natexlab{b}}, \apjl, 788, L13, \dodoi{10.1088/2041-8205/788/1/L13}

\bibitem[{{ZuHone} {et~al.}(2016){ZuHone}, {Markevitch}, \& {Zhuravleva}}]{zuhone2016mapping}
{ZuHone}, J.~A., {Markevitch}, M., \& {Zhuravleva}, I. 2016, \apj, 817, 110, \dodoi{10.3847/0004-637X/817/2/110}

\end{thebibliography}
\bibliographystyle{aasjournal}



\end{document}